\documentclass[12pt]{article}
\usepackage{blindtext, rotating}
\usepackage[english]{babel}
\usepackage{latexsym,amsmath,amssymb,bbm,amsfonts}
\usepackage[nohead]{geometry}
\usepackage[onehalfspacing]{setspace}
\usepackage[bottom]{footmisc}
\usepackage{caption}
\usepackage{subcaption}
\usepackage{graphicx}
\usepackage{bbding}
\usepackage{pifont}
\newcommand{\xmark}{\ding{55}}%
\usepackage{array} 
\usepackage[round]{natbib}
\usepackage{xcolor}
\usepackage[flushleft]{threeparttable}
\usepackage{rotating}
\usepackage{adjustbox}
\begin{document}

\begin{titlepage}

\singlespacing
\title{\vspace{-15mm}Measuring the Impact of Taxes and Public Services on Property Values: A Double
Machine Learning Approach}

\author{Anna Grodecka-Messi\thanks{Sveriges Riksbank, e-mail: anna.grodecka.messi@riksbank.se.}  \hspace{5mm} Isaiah Hull\thanks{Correspondence Address: Research Division, Sveriges Riksbank, SE-103 37, Stockholm, Sweden. Email: isaiah.hull@riksbank.se. Tel: +46 076 589 0661. Fax: +46 8 0821 05 31}}

\date{\today}

 \maketitle

\begin{center}
\textbf{Abstract}\\[0.05 cm]
 \end{center}
\linespread{1.00}
\small
How do property prices respond to changes in local taxes and local public services? Attempts to measure this, starting with \cite{Oates1969}, have suffered from a lack of local public service controls. Recent work attempts to overcome such data limitations through the use of quasi-experimental methods. We revisit this fundamental problem, but adopt a different empirical strategy that pairs the double machine learning estimator of \cite{Chernozhukov18} with a novel dataset of 947 time-varying local characteristic and public service controls for all municipalities in Sweden over the 2010-2016 period. We find that properly controlling for local public service and characteristic controls more than doubles the estimated impact of local income taxes on house prices. We also exploit the unique features of our dataset to demonstrate that tax capitalization is stronger in areas with greater municipal competition, providing support for a core implication of the Tiebout hypothesis. Finally, we measure the impact of public services, education, and crime on house prices and the effect of local taxes on migration.
\textbf{JEL-Classification}: C45, C55, H31, H41, R3\\

\end{titlepage}
\section{Introduction}

\noindent The choice and acquisition of housing is often one of the most important financial decisions an individual will make over the course of her lifetime. Beyond providing direct utility, housing services are associated with a range of local public goods that are provided on a neighborhood or district level. As such, the price of housing is not only a function of its physical attributes, but also the characteristics of its surroundings and the access it provides to local services. Taxes levied within a region or community to pay for such services will also affect local property prices and the extent to which they do has been the subject of a vast literature on tax capitalization. Much of the early work on this subject, however, suffered from severe bias in estimates of tax capitalization due to the omission of important local controls and also failed to adequately account for the endogeneity of local taxes (\citealp{Wales74}; \citealp{Palmon1998}; and \citealp{Ross1999}).

Modern work on tax capitalization improves on the deficiencies of the early literature by making use of the boundary discontinuity design introduced by \citet{Black99} and by employing quasi-experimental methods\footnote{See \citet{Bradley2017} and \citet{Oliviero19}.} to circumvent the need for local controls. While these recent efforts improve on earlier work and provide estimates with a causal interpretation, they typically are not able to comment on the importance of local public services that lie at  the heart of the \cite{Tiebout56} hypothesis, which is closely tied to the optimal provision of public goods. Furthermore, the strong association with a single event -- typically a legal or regulatory reform -- weakens the case for external validity. 

Much of the recent literature also struggles to convincingly resolve the tax endogeneity issue, since it cannot account for the underlying drivers of tax changes and does not have access to local political or economic controls. As \cite{Semenova2020} argue, there are two main approaches that can be used to solve the endogenous regressor problem: account for the sources of endogeneity or find an instrumental variable. Our paper follows the first approach by pairing the ``double machine learning'' method developed by \citet{Chernozhukov17, Chernozhukov18} with a unique and exhaustive dataset of 947 time-varying, municipal-level controls in Sweden for the 2010-2016 period. The breadth of our dataset enables us to remove the impact of a wide variety of confounders -- including the political and economic drivers of tax changes -- yielding a measure of the treatment variable that can be viewed as randomly assigned. Consequently, our results have a causal interpretation under the assumption of unconfoundedness.

Our main empirical exercises estimate the impact of taxes, public service inputs, public service outputs, and crime on house prices. We also measure the impact of taxes on migration to test the mechanism underlying the Tiebout hypothesis. Apart from accounting for the time-varying local characteristics and public service controls, the data also enables us to include municipality and year fixed effects to control for any remaining unobservable differences over geography or time. Importantly, our approach also produces results that are more likely to have external validity than work in the existing literature, since we do not rely on a specific experiment or a legal or regulatory change in a single region. As such, our approach can be replicated in other countries, conditional on data availability. Furthermore, given the strong trend toward the cultivation of ``big data'' sources and the increased adoption of machine learning methods in economics, we anticipate that the approach in this paper will have increasing relevance for future work.

While our dataset is novel and our estimation methods are built around recent work in econometrics \citep{Chernozhukov17, Chernozhukov18} and machine learning \citep{Cheng16}, the core question we evaluate in this paper has been present in the economic literature for decades. In his seminal paper, \cite{Tiebout56} argued that households, while choosing where to live, ``vote with their feet," revealing their preference for a mix of local public goods and taxes. He sketched out a theory that suggested that this act of preference revelation was sufficient to solve the free-rider problem for local public goods.\footnote{Prior to \cite{Tiebout56}, \cite{Musgrave1939} and \cite{Samuelson1954} had developed a theory of public goods provision, but did so with respect to aggregate public goods, and found that no market-type mechanism could determine the efficient level of provision.} \cite{Oates1969} provided the first empirical test of the Tiebout hypothesis (1956) and linked property values in the community to the local public budgets, focusing on the effects of property taxation and local expenditure on housing values. He documented that approximately two thirds of changes in property taxation are capitalized into property values. Furthermore, \cite{Oates1969} also started a vast empirical and theoretical literature on the capitalization of local taxes into housing values, which has focused primarily on property taxes.\footnote{See \cite{Pollakowski73}, \cite{Edel74}, \cite{Hamilton76}, \cite{Meadows76}, \cite{King77}, \cite{Rosen77}, \cite{Epple78}, \cite{Cebula78}, \cite{Brueckner79}, \cite{Reinhard81}, \cite{Goldstein81}, \cite{Yinger82}, \cite{Rosen82}, \cite{Mieszkowski89}, \cite{Palmon1998}, \cite{Bai14}, and \cite{Elinder2017}.} 

We contribute to this literature by providing unbiased estimates of tax and public service capitalization, and by exploiting our unique dataset to provide explicit tests of the Tiebout hypothesis and its core implications. Given that the property tax is set at the national level in Sweden, we focus on income taxes instead, which vary at the municipal level. While local income taxes are less common than local property taxes, they are also used in several other countries, including Switzerland, Finland, and Denmark (\citealp{Kitchen2004}). An earlier study for Sweden \citep{Boije} provides income tax capitalization estimates for the Stockholm region and finds evidence for partial tax capitalization.\footnote{The tax capitalization literature mostly focuses on estimating property tax capitalization, since most of the work uses U.S. data, where local variation in taxes comes primarily from property taxes; however, a number of papers also focus on capitalization of local income taxes (\citealp{Rosen79}; \citealp{Stull1991}; \citealp{Morger2017}; \citealp{Basten2017}). The last two papers focus on Swiss data. Switzerland grants its municipalities substantial autonomy in determining their finances, and being one of the most prominent examples of fiscal federalism, it provides for a good testing ground for the tax capitalization effect. However, as \cite{Morger2017}, p. 247, notes ``for Switzerland, information on the quantity and quality of public goods provided at the local level is nonexistent."} 

Our main exercise estimates local income tax capitalization into house prices in the presence of an exhaustive dataset on local public services and local characteristics. We first use an OLS specification that includes municipal fixed effects, annual fixed effects, and the set of time-varying controls that is most commonly used in the literature. We find that a one standard deviation increase in local income taxes is associated with a 0.13 standard deviation decrease in property prices. We then apply the double machine learning (DML) estimator \citep{Chernozhukov17, Chernozhukov18}, coupled with a deep-wide network architecture \citep{Cheng16} to estimate tax capitalization in the presence of a high dimensional and potentially nonlinear nuisance parameter. This approach allows for both the use of fixed effects, which enter the model linearly, and a flexible and nonlinear specification for the local public service and characteristic controls. Within this framework (hereafter, DML-DW), we find that a one standard deviation increase in local income taxes decreases property prices by 0.26 standard deviations. As such, proper accounting for public services more than doubles our estimate of the tax capitalization effect by reducing the downward bias in magnitude (\citealp{Palmon1998}, p. 1108). Beyond quantifying the size of the bias, we are also able to demonstrate that it appears to arise primarily from the omission of housing variables (including supply), public finance measures, and public service outputs. This has problematic implications for estimates in the literature, including otherwise well-identified estimates, since the tax changes themselves are likely to depend on such variables.

We also show how this new approach for estimating tax capitalization can be used to test different aspects of the \cite{Tiebout56} model. Given the \cite{Tiebout56} theory's prediction that voting with one's feet depends on households being highly mobile and having low moving costs, our tax capitalization results should be stronger in densely populated counties with many municipalities, where households can move at a low cost and without changing jobs. Our results confirm this claim: In urban areas, a one standard deviation increase in the municipal income tax reduces property prices by 1.04 standard deviations. In rural areas, where people's choice of municipalities is limited, the tax capitalization result is almost non-existent: A one standard deviation increase in income taxes decreases property prices by a mere 0.01 standard deviations. Following \cite{Cebula78} and \cite{Banzhaf08}, we also test whether people indeed move in response to higher income taxes and document that a one standard increase in income taxes has a small, but negative impact of net migration. Again, this effect is almost twice as large for high density counties with many municipalities. 

Finally, our broad set of controls and empirical framework can be jointly used to test for the capitalization of different public services into property prices and for examining differential impacts of public service inputs and outputs on house prices. A large literature studies the impact of schooling on house prices (\citealp{Haurin96}; \citealp{Black99}; \citealp{Downes2002}; \citealp{Barrow04}; \citealp{Cheshire04}; \citealp*{Bayer07}; \citealp{Ries2010}). We show that once public services are controlled for properly, schooling inputs (expenditure) have a negative impact on house prices and outputs (test scores and other quality measures) typically have a positive impact. The literature generally finds a positive association between schooling expenditures and house prices; however, this is likely because schooling outputs, such as grades, are correlated with schooling inputs. Using DML-DW and controlling exhaustively for public services, including schooling outputs, we find a negative association between spending per pupil and house prices, which suggests that the positive association documented in the literature is likely to be driven by its relationship with outputs. This supports the sub-literature that emphasizes the use of public service outputs, rather than inputs (\citealp{Rosen77}; \citealp{Hanushek86}; and \citealp{Hanushek96}).\footnote{Note that \cite{Tiebout56} writes about local public expenditures. However, \cite{Oates1969} already admits that having a measure of public service outputs would be ideal for testing the Tiebout hypothesis. He chooses to work with a measurement of expenditure per pupil due to data availability, but considers it an imperfect variable for the purpose of the test.} 

We also contribute to a closely related literature, which examines the effect of crime rates on property prices and expenditures, documenting a negative relation between the two (\citealp{Thaler78}; \citealp{Reinhard81}; \citealp{Blomquist88}; \citealp{Haurin96}; \citealp{Gibbons04}; \citealp{Linden08}). In line with the literature, we also find a negative relationship between crime and house prices. Again, we find that his effect is larger in urban areas, where municipal competition is higher.

The paper proceeds as follows. We describe the data used in Section \ref{Data}. We proceed with the estimation strategy in Section \ref{Est}. Section \ref{Res} presents and discusses the empirical results. Finally, we conclude in Section \ref{Con}.
\section{Data}\label{Data}

\begin{figure}[ht!]
  \caption{Frequency of municipal-level control variables by type}
  \includegraphics[width=\linewidth]{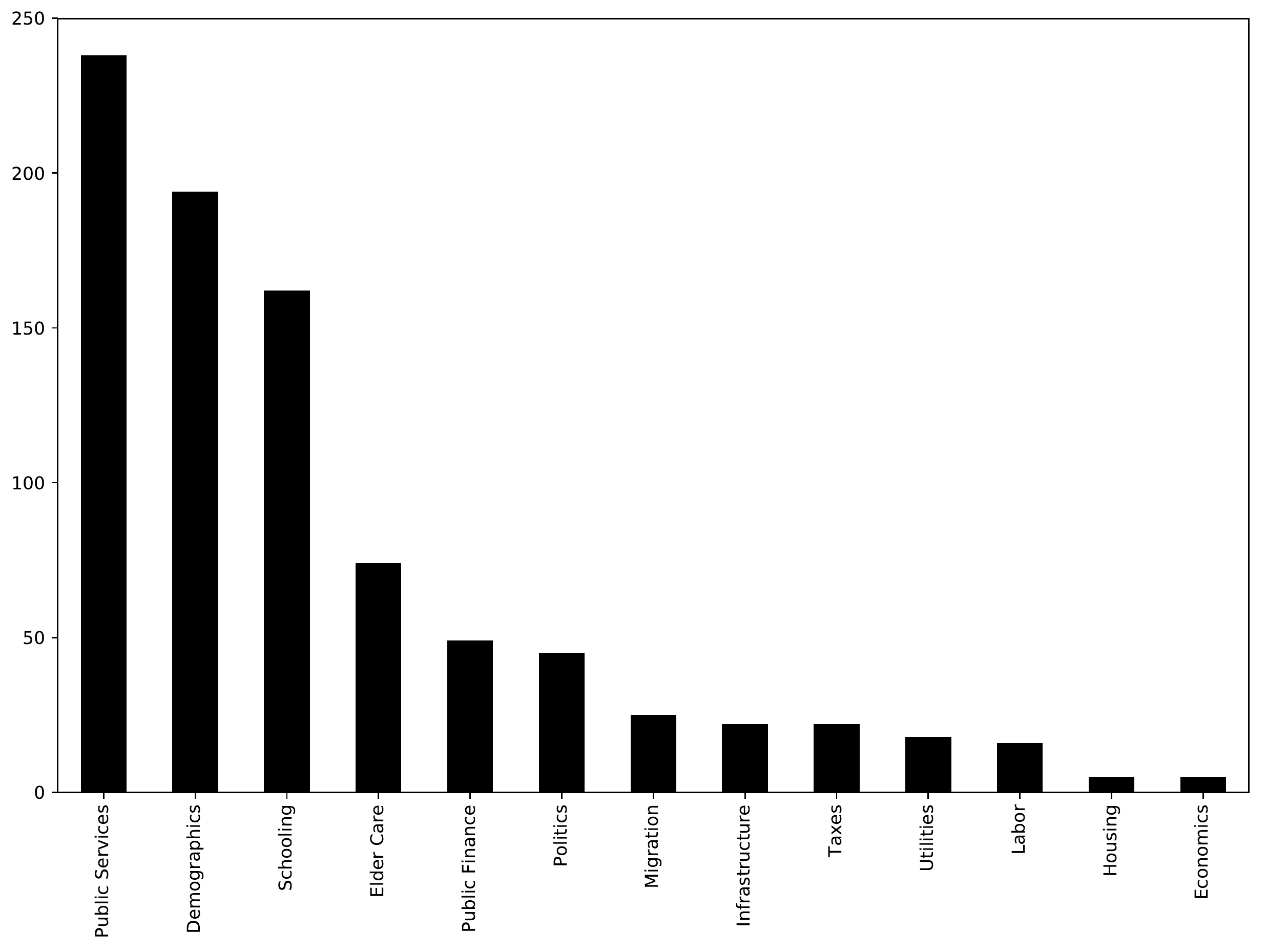}
  \label{fig:controls_hist}
\footnotesize
$Notes$: Our dataset contains 947 municipal-level control variables. This spans categories from demographics and migration to public service inputs and outputs. We assembled the dataset using webscraping and then manually classified the variables into categories using variable titles and descriptions.
\end{figure}

We combine three main data sources: data on local house prices, announcements of changes in local income taxes, and a database of public good provision and municipal characteristics. The housing data is scraped from an exhaustive online source of housing transactions.\footnote{The microdata is collected from booli.se, which aggregates housing transactions. Note that we do not perform hedonic adjustments to the square meter price prior to its inclusion as the dependent variable in regressions; rather, our specifications include controls that capture local amenities and characteristics, as well as changes in the housing stock within the sample period. We also include geographic fixed effects that capture the initial composition of the local housing stock. We do not find any substantive differences in our results if we instead use official house price series with hedonic adjustments; however, this reduces our sample size by limiting the number of municipalities that can be included.} Yearly announcements in local income taxes are provided by Statistics Sweden. The public good provision and municipal characteristics data is scraped from an online aggregator of local and regional statistics.\footnote{We scraped the control variables from kolada.se, which is an online aggregator of local and regional statistics for Sweden.} The data spans the 2010-2016 period and includes 947 potential covariates. Figure \ref{fig:controls_hist} shows the number of categories of series that are available by type. A finer disaggregation of the series types is given in the Appendix in Section \ref{app_series}. Note that we have standard controls, including local economic, housing, and labor market conditions. We also have less commonly used controls, such as local public services, demographics, schooling, politics, infrastructure, migration, and public finance. Controlling for local political outcomes and the state of local public finance enables us to credibly deal with the tax endogeneity issue.

Part of Sweden's system of public finance is centralized and conducted at the national level, but other parts of the tax system and many public services are decentralized. For example, services such as education, childcare, healthcare, and elderly care are provided at the municipal level and are financed by local income taxes. Since 2005, there has been a system in place that redistributes taxes between the wealthiest and the poorest municipalities in order to provide a basic level of services across the country (\citealp{hansson2020}).\footnote{There are different types of transfers that a municipality can receive: For example, the income redistribution system collects a fee from the twelve municipalities with the highest tax revenue and transfers the proceeds to the remaining municipalities in the form of a benefit (\citealp{SKR}).} Despite the system of redistribution, there is still variation in the access to and quality of certain services across municipalities. Income taxes, which lie at the heart of our study, constitute the largest source of tax revenue in Sweden.

Given the decentralization of the public system, we conduct our analysis at the municipality level. Sweden is divided into 21 counties and 290 municipalities, which have discretion over the determination of local income taxes.\footnote{One county, Gotland, is also a municipality. Consequently, it exclusively has a municipal tax, which is higher than for other counties. It is also an island and, thus, has unusually low municipal competition. Note that our results are not sensitive to removing it from the sample.} While income taxes in Sweden are progressive, the municipal component is not. Rather, it is identical for all individuals and is proportional to income. Above a certain income threshold, individuals also pay a federal tax on marginal earnings. It is this additional federal tax that drives the progressiveness of the system. Importantly, local and federal income taxes do not interact in any way. All individuals have to pay the local income tax and high earners pay an additional federal tax. Neither the income threshold for the payment of the federal tax nor the rate of the federal income tax are dependent on the level of local income taxation and the local tax paid by the citizen. It is thus the variation in income taxes levied on municipal level that may drive differences in house prices across municipalities. As of 2018, it is estimated that the incomes of 64\% of people working in Sweden were only taxed at the municipal level (\citealp{Lidefelt2017}). In past years, this percentage has exceeded 70\%, indicating that the level of income taxes is the most important determinant of the after-tax income of the majority of individuals.

\begin{figure}[ht!]
 \caption{Geographical distribution of municipal taxes in Sweden}
    \begin{center}
    \begin{subfigure}[t]{0.49\textwidth}
        \begin{center}
        \includegraphics[width=5cm,trim={0cm 0cm 0 0}, clip]{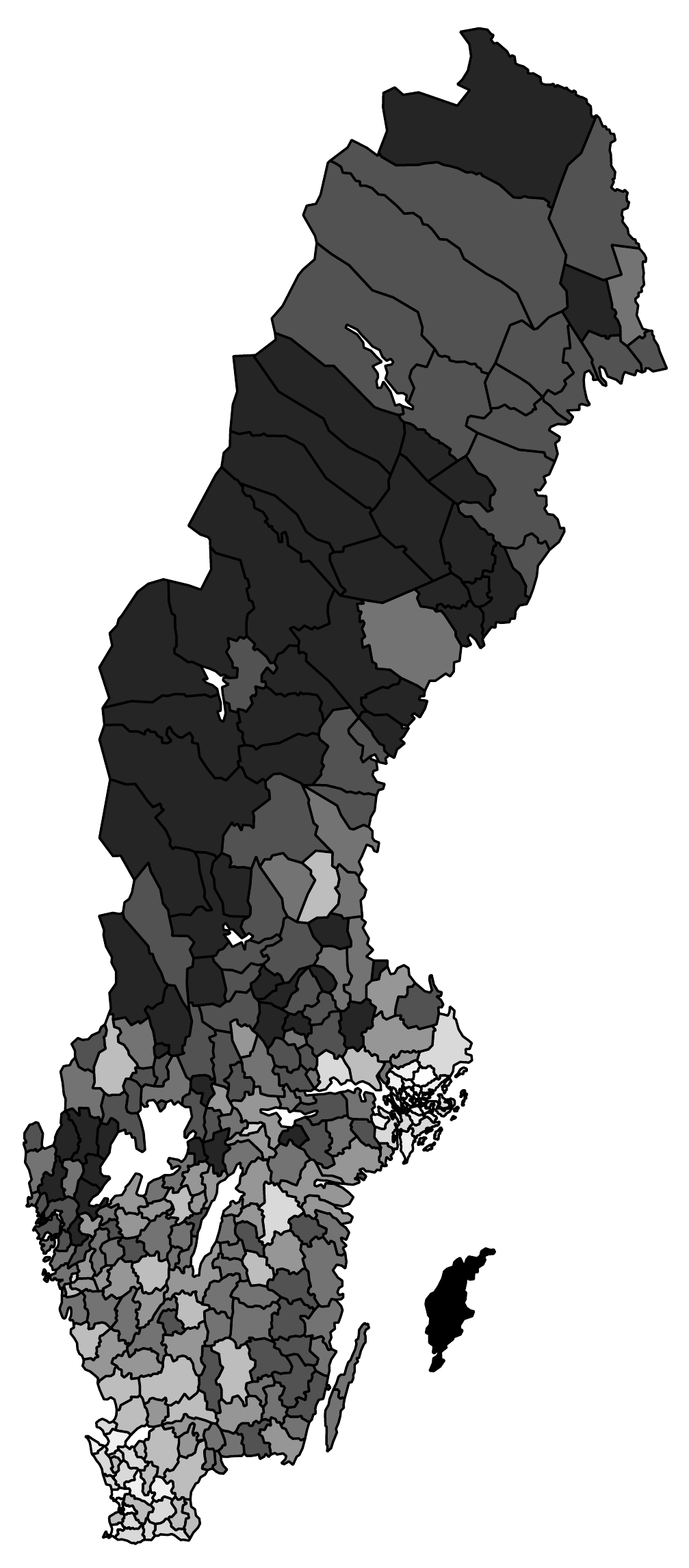} 
        \caption{Municipal Tax Level: 2010} \label{fig:municipal_taxes_2010}
        \end{center}
    \end{subfigure}
    \hfill
    \begin{subfigure}[t]{0.49\textwidth}
        \begin{center}
        \includegraphics[width=5cm,trim={0cm 0cm 0 0}, clip]{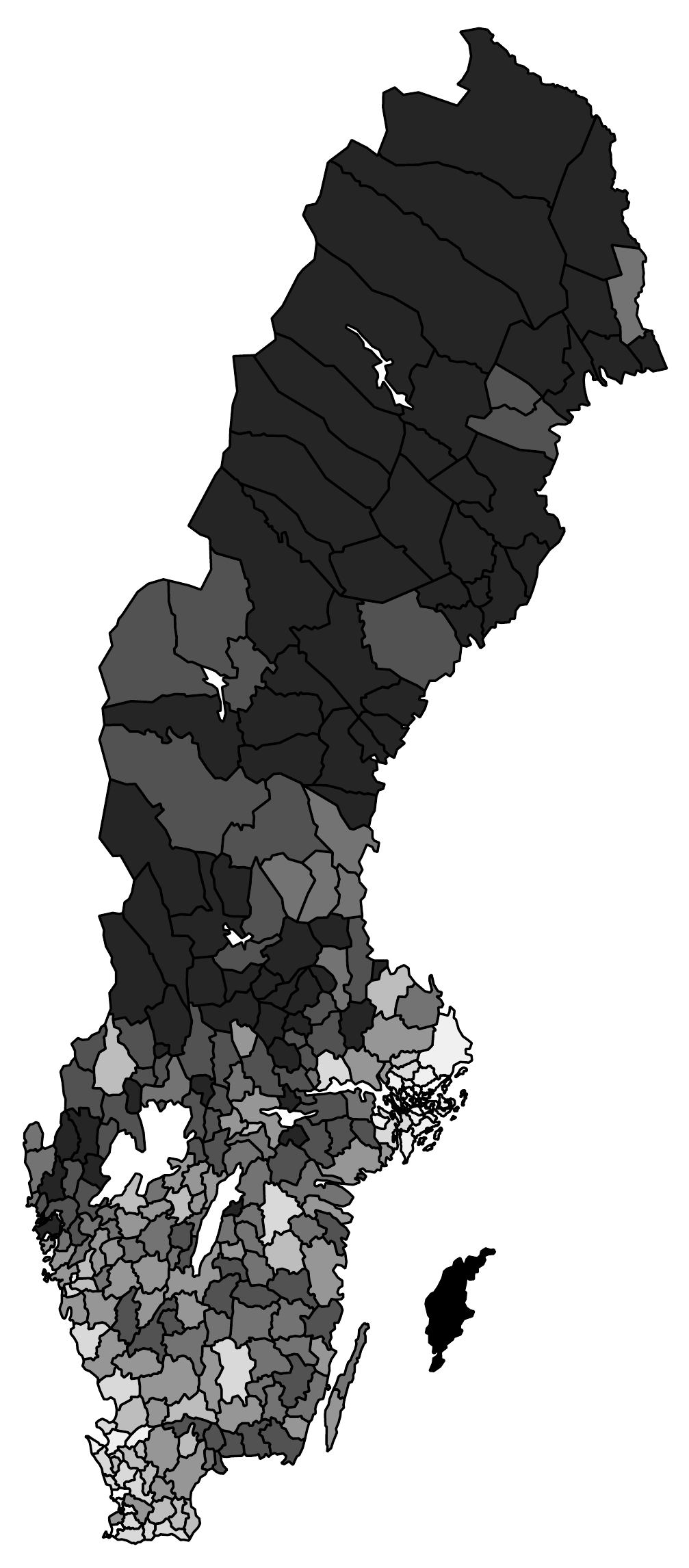}
        \caption{Municipal Tax Level: 2016} \label{fig:municipal_taxes_2016}
        \end{center}
    \end{subfigure}
   \end{center}
\footnotesize
$Notes$: The subfigures above show tax rates in the 290 Swedish municipalities. A darker shade indicates a higher rate. Note that subfigure (a) shows the tax rates in 2010 and subfigure (b) shows the tax rates in 2016. Large and geographically isolated municipalities often have higher rates.
\label{fig:tax_maps}
\end{figure}

The density of municipalities is highest in the south of Sweden and around the largest cities. For example, Stockholm County is divided into 27 municipalities with different local tax values. Municipality boundaries sometimes run through the middle of the street, so that it suffices to move from one side of the street to the other in order to face a much lower tax burden.\footnote{It is important to stress that, in Sweden, property taxes are set on a national level. A study testing the implications of a change in property taxes found that the effect of a decrease in property tax has approximately no effect for most properties (\citealp{Elinder2017}).} Figure \ref{fig:tax_maps} shows municipal tax levels for 2010 and 2016, with darker shades indicating higher income tax rates. Notice that southern Sweden is much more densely populated than its northern part, so we might expect tax capitalization to have a stronger effect there. We specifically test for this in Section \ref{Rex_tax}.

\begin{figure}[ht!]
  \caption{The municipal tax distribution for selected years}
  \includegraphics[width=\linewidth]{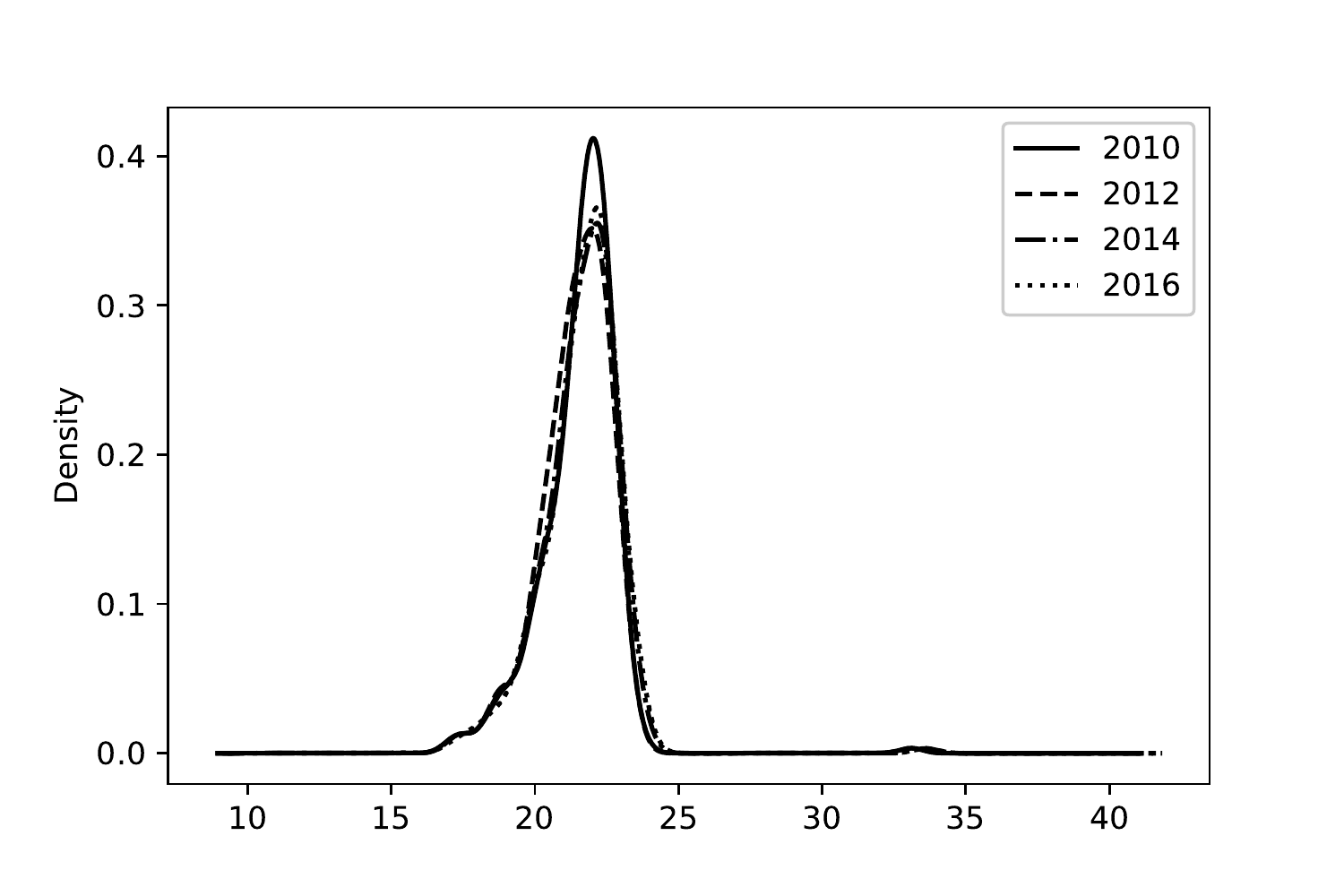}
  \label{fig:taxes}
\footnotesize
$Notes$: The figure above shows kernel density plots of the tax distributions for selected years: 2010, 2012, 2014, and 2016. While the plots display tax rates for every second year, we use annual tax rates in the regression analysis. Note that most rates are concentrated between 15\% and 25\%. Additionally, between 2010 and 2016, we observe a slight rightward shift in the distribution of taxes.
\end{figure}

The distribution of taxes across municipalities is shown in Figure \ref{fig:taxes}. Each plot displays the kernel density for a given year. While our set of controls is restricted to the years 2010-2016, tax data is available between 2001 and 2018. We show one plot for every two years, starting in 2010 and ending in 2016. Over this period, the distribution shifted to the right, but only slightly, suggesting that taxes tended to increase on average.

\begin{figure}[ht!]
  \caption{The municipal tax change distribution for selected years}
  \includegraphics[width=\linewidth]{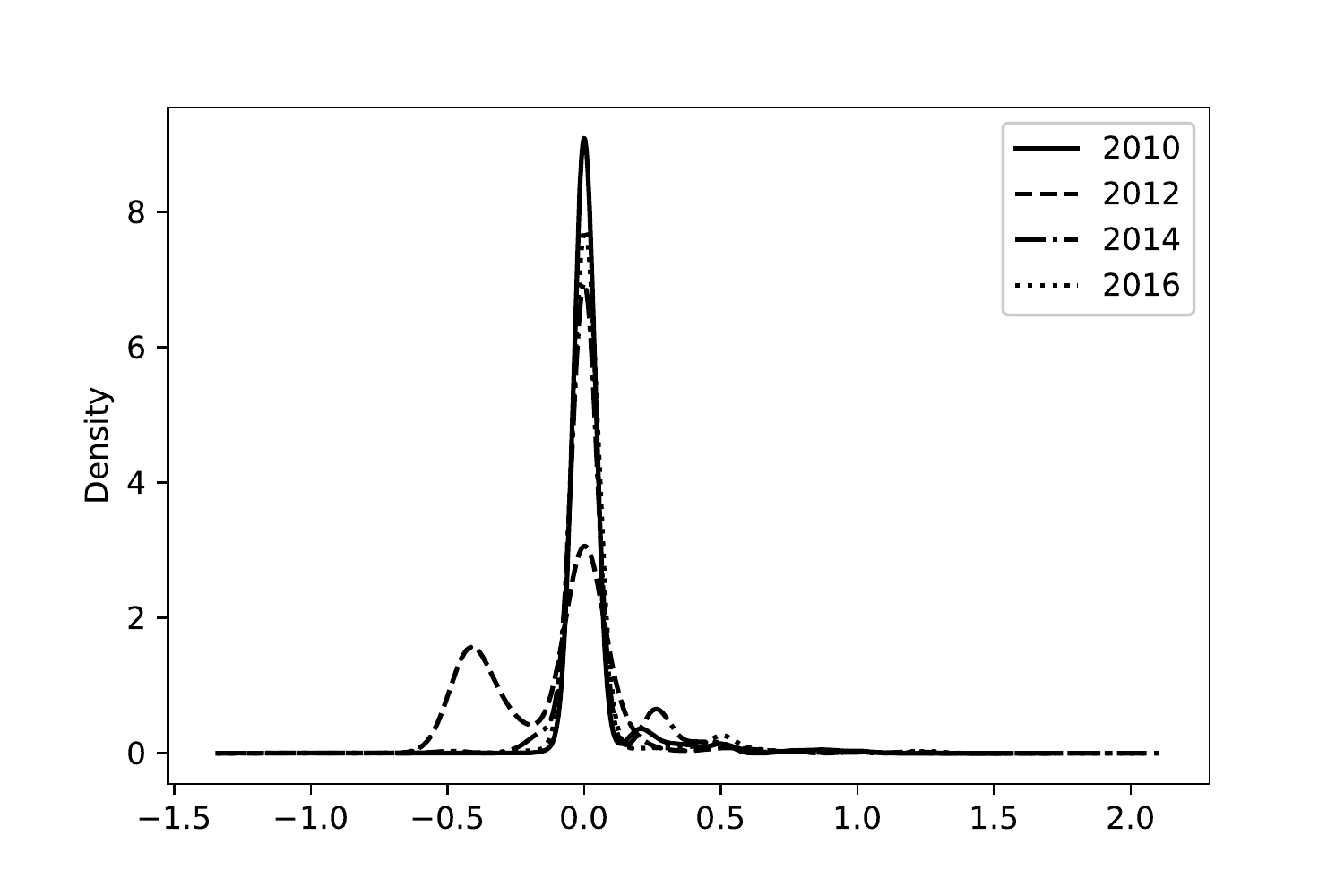}
  \label{fig:taxdiff}
\footnotesize
$Notes$: The figure above shows the distribution of municipal-level tax changes in selected years. The tax rate change for 2010, for instance, is computed as the rate that applied in 2010, less the 2009 rate. Note that there are substantial differences across years in the share of municipalities with tax rate changes.
\end{figure}

A rightward shift in the tax distribution does not imply that all municipalities tended to see an increase in taxes. In Figure \ref{fig:taxdiff}, we can see the distribution of all municipal tax changes within a given year. We have both tax increases and decreases in all years. There are also differences in how numerous tax changes are across years. For instance, 2010 witnessed relatively few tax changes in comparison to 2012. 

\begin{figure}[ht!]
 \caption{Municipal-Level House Price Growth (Year-over-Year)}
    \begin{center}
    \begin{subfigure}[t]{0.49\textwidth}
        \begin{center}
        \includegraphics[width=7.5cm,trim={0cm 0cm 0 0}, clip]{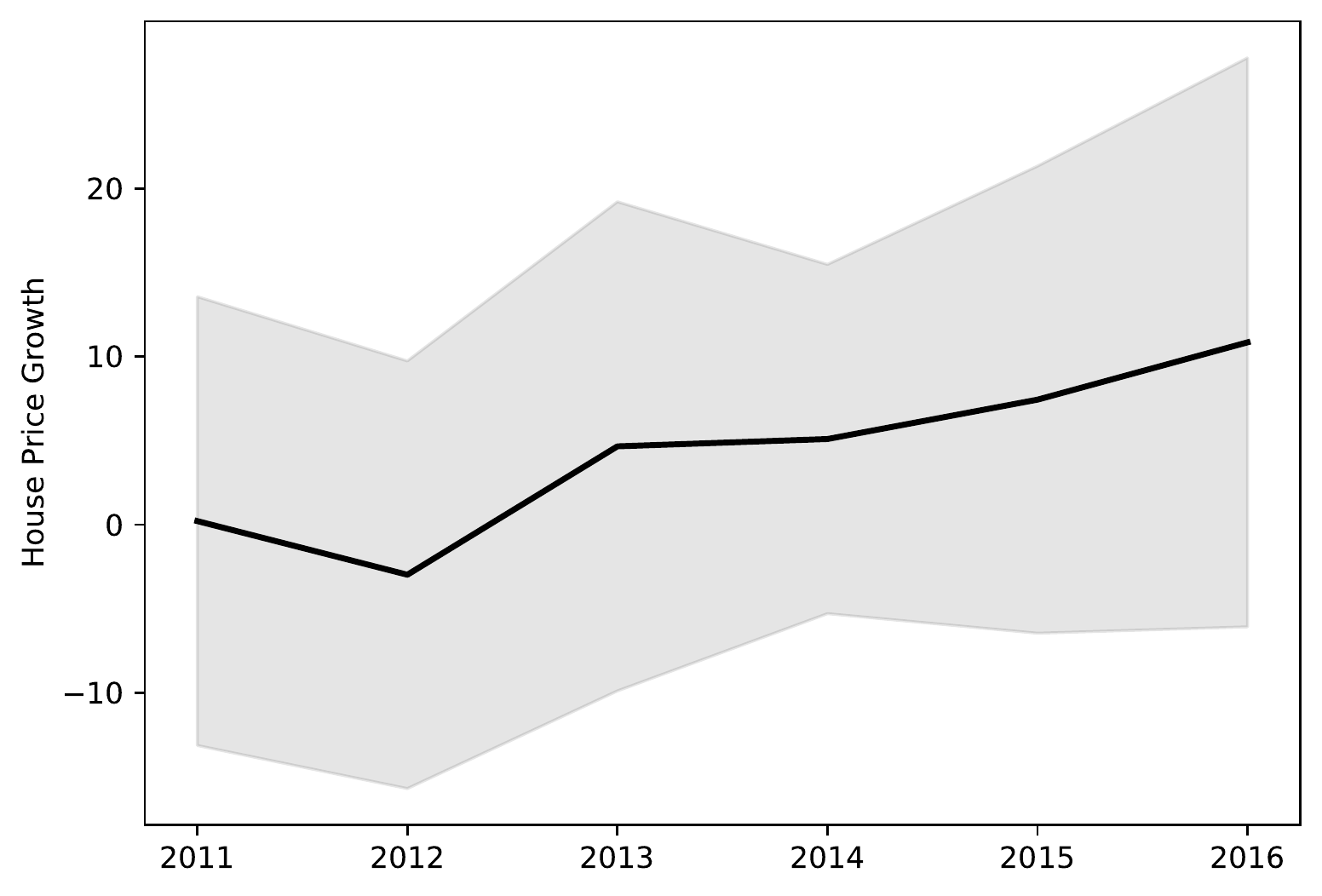} 
        \caption{Mean Price Growth} \label{fig:time_series_hp}
        \end{center}
    \end{subfigure}
    \hfill
    \begin{subfigure}[t]{0.49\textwidth}
        \begin{center}
        \includegraphics[width=7.5cm,trim={0cm 0cm 0 0}, clip]{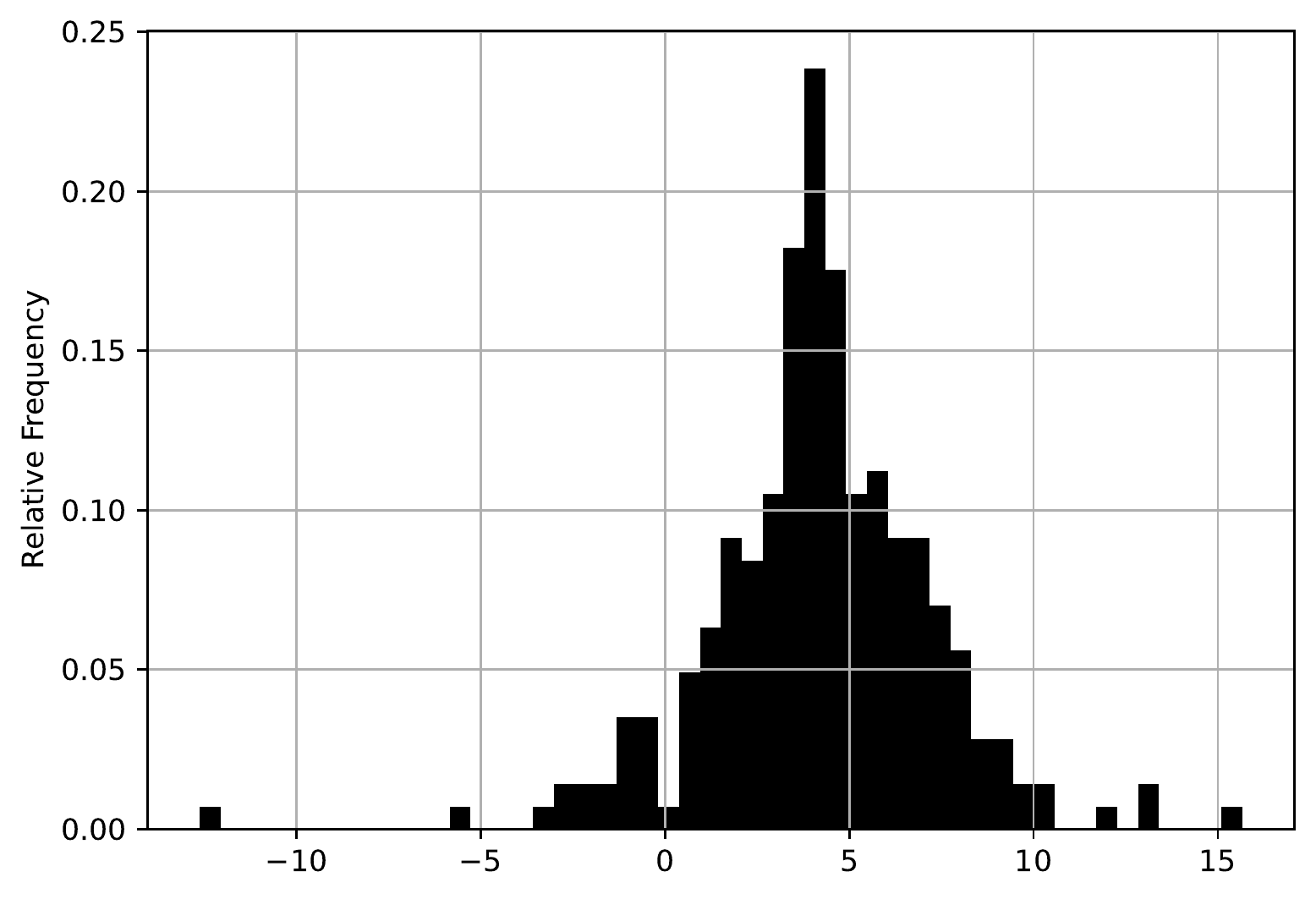}
        \caption{Cross Sectional Dispersion} \label{fig:histogram_hp}
        \end{center}
    \end{subfigure}
   \end{center}
\footnotesize
$Notes$: The subfigures above show year-over-year square meter house price growth at the municipal level. Subfigure (a) plots the mean growth rate across municipalities, along with a one standard deviation confidence band. Subfigure (b) shows a histogram of year-over-year growth rates for each municipal-year.
\label{fig:house_price_plots}
\end{figure}

With respect to property prices, Sweden underwent a long period of increasing housing prices, and on average, experienced price increases in most municipalities over our sample period. Figure \ref{fig:time_series_hp} shows the mean municipal house price growth rate, along with one standard deviation confidence bands. While growth was uniformly strong, there were still considerable differences across municipalities, as indicated by Figure \ref{fig:histogram_hp}, which presents a histogram of municipal square meter house price growth rates.

Housing prices in Sweden are a good indicator of property values. The majority of households are homeowners and the rental markets are regulated, so rental prices set centrally below market values are not a good indicator of developments in the housing market (\citealp{Englund2011}). There are big differences in house prices among different regions of Sweden, depending on the density of urbanisation and housing shortage (\citealp{emanuelsson2015}). The summary statistics presented in Table \ref{tab:descriptive_statistics} suggest that there are stark differences between house prices in rural and urban areas.

Table \ref{tab:descriptive_statistics} also provides summary statistics for all regressors of interest and dependent variables. In the top panel, descriptive statistics are shown for all 290 municipalities over the 2010-2016 period. They are also given for the urban and rural municipalities separately in the lower panels. The municipal tax is given as a rate. The square meter house price is computed as the mean price of a single family villa and is given in thousands of Swedish kronor (SEK).\footnote{We use a novel dataset of housing transactions in Sweden to compute the mean square meter house price at the municipal level. The summary statistics for our data are consistent with commonly-used private indices, which use similar transaction-level data to construct indices.} Net migration is computed as the total number of immigrants minus the total number of emigrants at the municipal level. Schooling expenditures are computed as the net cost of upper secondary school in thousands of SEK per student. Grades are computed as an average at the municipal level, where A = 20, B = 17.5, C = 15, D = 12.5, E = 10, and F = 0. And crime is the total number of reported violent crimes in the municipality per 100,000 residents. Note that the variables are standardized prior to estimation to eliminate scaling issues.

\begin{table}[hb!]
\caption{Descriptive statistics for regressors of interest and dependent variable}
\resizebox{\linewidth}{!}{
\centering
\begin{tabular}{lcccccc}
\hline
\hline 
\textbf{All Municipalities} & Tax & House Prices &  Migration & Schooling Costs & Grades &  Crime \\
\hline
Mean & 0.22 & 13.74 & 120.76 & 113.45 & 14.55 & 895.04 \\
Std & 0.01 & 11.04 & 333.85 & 15.58 & 0.53 & 320.61 \\
Min & 0.17 & 1.54 & -173 & 64.76 & 9.555 & 136.61 \\
25\%	 & 0.21 & 6.96 & 4 & 103.17 & 14.26 & 685.29 \\
50\%	 & 0.22 & 10.00 & 48 & 111.78 & 14.56 & 867.97 \\
75\% & 0.22 & 17.21 & 126 & 120.00 & 14.84 & 1069.18 \\
Max & 0.34 & 152.66 & 5742 & 216.70 & 16.88 & 3647.85 \\
\hline
N & 1764 & 1764 & 1764 & 1764 & 1764 & 1764 \\
\hline 
\textbf{Rural} & Tax & House Prices &  Migration & Schooling Costs & Grades &  Crime \\
\hline
Mean & 0.22 & 9.93 & 78.58 & 116.83 & 14.49 & 887.49 \\
Std & 0.01 & 5.47 & 145.82 & 16.48 & 0.51 & 300.77 \\
Min & 0.2	& 1.54 & -163 & 64.76 & 9.56 & 136.61 \\
25\%	 & 0.22 & 6.18 & -3 & 105.77 & 14.22 & 696.81 \\
50\%  & 0.22 & 8.49 & 34 & 114.47 & 14.54 & 870.13 \\
75\%  & 0.22 & 12.6 & 103 & 124.87 & 14.82 & 1060.67 \\
Max 	& 0.34 & 36.59 & 1633 & 216.7 & 16.1 & 3022.93 \\
\hline 
N & 1106 & 1106	& 1106 & 1106 & 1106 & 1106 \\
\hline 
\textbf{Urban} & Tax & House Prices &  Migration & Schooling Costs & Grades &  Crime \\
\hline
Mean & 0.21 & 20.15 & 191.65 & 107.77 & 14.63 & 907.73 \\
Std & 0.01	 & 14.54 & 505.27 & 11.97 & 0.55 & 351.31 \\
Min& 0.17	 & 4.67 & -173 & 75.36 & 12.92 & 208.75 \\
25\%	 & 0.2 & 9.27 & 20 & 99.39 & 14.31 & 660.18 \\
50\%	 & 0.21 & 17.09	 & 80 & 107.05 & 14.59 & 857.68 \\
75\%	 & 0.22 & 26.4 & 171	 & 115.09 & 14.93 & 1087.75 \\
Max	& 0.24 & 152.66 & 5742 & 149.97 & 16.88	 & 3647.85 \\
\hline
N & 658 & 658 & 658 & 658 & 658 & 658 \\
\hline 
\end{tabular}}
\scriptsize
$Notes$: Descriptive statistics are computed for all municipalities, urban municipalities, and rural municipalities separately. The sample includes coverage for 290 municipalities over the 2010-2019 period. The municipal tax is given as a rate. House prices are computed using transactions of single family houses and are given in thousands of Swedish kronor (SEK) per square meter. Net migration is computed as the number of immigrants minus the number of emigrants. Schooling expenditures are computed as the net cost of upper secondary school in thousands of SEK per student. Grades are computed as an average at the municipal level, where A = 20, B = 17.5, C = 15, D = 12.5, E = 10, and F = 0. Violent crime is the number of reported violent crimes in the municipality per 100,000 residents.
\label{tab:descriptive_statistics}
\end{table}

\section{Estimation Strategy}\label{Est}

In this section, we describe our empirical strategy for measuring tax and public service capitalization into house prices. We start by describing the tax capitalization estimation problem and how existing work in the literature has dealt with it. After that, we discuss the estimation technique we use, which may be useful for empirical work in settings where the econometrician has a high number of time-varying covariates and fixed effects.

\subsection{The Estimation Problem}\label{Est_problem}

Tax capitalization estimates in the early literature suffered from a lack of availability of high quality public service controls. Most papers, starting with \cite{Oates1969}, which used public school expenditure per pupil in its main regression specification, had at most two public services as controls (\citealp{Pollakowski73}; \citealp{King77}; \citealp{Rosen77}; \citealp{Cebula78}; and \citealp{Brueckner79}). Since higher taxes are likely to be associated with higher public service provision, omitting them from the regression specification will lead to biased estimates of tax capitalization. Furthermore, if local taxes tend to capitalize negatively into house prices and local public service provision tends to capitalize positively, the direction of the bias will be positive. If the true effect of tax capitalization is negative, then the positive bias will reduce the estimated magnitude.

Early empirical work tried to improve on Oates' original regressions by including more public services in the regression specification. \citet{Reinhard81}, for instance, included recreational expenditures per capita, crime rates, and expenditures on streets and highways, contrasting with much of the literature, which used only one measure of public expenditures. Still, as \cite{Palmon1998} note, the literature following \cite{Oates1969} made only incremental improvements over the years and the 
omission of local public service controls remained a problem due to data availability. \cite{Palmon1998} overcame the potential bias in the tax capitalization estimates created by an inadequate measurement of public services (or a lack thereof) and the comovement between public services and local taxes by focusing on local areas where public services were essentially fixed, while local taxes varied. 

In addition to the lack of local public service controls, the empirical literature has also suffered from limitations in time and geographic variation. Most work has examined a small number of municipalities in a particular region and year. Consequently, such estimates of tax capitalization were identified purely off of cross-sectional variation. Given the lack of geographic fixed effects, these estimates were likely to have confounded tax capitalization with permanent features of geographic locations, such as the unchanging components of housing supply elasticity (e.g. \citealp{Saiz10}). If taxes tend to be higher in areas with low supply elasticities, such as coastal cities, then we might expect the omission of geographic fixed effects to create a positive bias in tax capitalization measurements. Overall, however, the impact of the omission of geographic fixed effects is directionally ambiguous. 

In addition to the inclusion of public service controls and adequate time and geographic variation, much of the empirical literature also lacks local characteristic controls. Demographics, migration, labor markets, political preferences, housing supply, and economic conditions all comove cross-sectionally and over time with both taxes and house prices. Even in a specification with sufficient time variation to include geographic fixed effects, biases will still emerge from failing to properly control for movements in local characteristics. As with the omission of geographic fixed effects, the direction of this bias is unclear ex-ante.

As \cite{Palmon1998} point out, such biases contaminated estimates in the empirical tax capitalization literature up until the late 1990s. More recent work has dealt with these biases by making use of better data and newer econometric methods. Notably, \cite{Basten2017} and \cite{Morger2017} make use of microdata on apartment rentals in Switzerland with substantial time and geographic variation. \cite{Basten2017} sets up a regression discontinuity design, exploiting the fact that income taxes depend discretely on residency, but public service access depends continuously on the distance from the service provided. \cite{Morger2017} uses a hedonic regression and exploits time and geographic variation to control for unobserved public services. In both cases, the authors use high quality rental data, coupled with substantial time and geographic variation, to obtain unbiased estimates of tax capitalization without including data on local public service provision, which is unavailable in Switzerland. Earlier work, including \cite{Black99}, made use of a border discontinuity design, but for the purpose of estimating the impact of public services on house prices. An alternative approach, which has recently been used in the literature, uses a quasi-experimental setup (see \citealp{Bradley2017} and \citealp{Oliviero19}) to overcome the aforementioned problems in the literature. Such work focuses on the effects of a specific one-off event that enables the establishment of causality.

We contribute to the literature by applying an alternative framework to estimate the tax capitalization effect, which does not rely on the assumption that local public services are fixed. The framework we use, which is not built around a specific change in law or regulation, makes use of an exhaustive set of local public services and local characteristics, as well as recently introduced work in econometrics \citep{Chernozhukov17, Chernozhukov18} and machine learning \citep{Cheng16}. Given the strong trend towards the cultivation and use of ``big data'' sources and the adoption of methods from machine learning in economics, it is likely that this framework will become increasingly relevant in future empirical work.

\subsection{Double Machine Learning}\label{Est_DML}

Our empirical strategy attempts to overcome the problems identified by the literature and, in doing so, makes two contributions. First, we introduce a novel and comprehensive dataset on house prices, public services, taxes, and local characteristics for Sweden. The dataset spans all 290 municipalities and covers the time period between 2010 and 2016. To our knowledge, it contains the most comprehensive set of public services and local controls of any paper in the tax capitalization literature. Second, we adopt an empirical strategy that is capable of handling a high number of covariates that may have a nonlinear relationship with the dependent variable. As \citet{Athey2017c} points out, this is precisely where machine learning (ML) algorithms are most useful in economics. They allow for a flexible functional form assumption, but require the researcher to impose discipline through the use of out-of-sample prediction. The standard model training process involves splitting the sample into training and validation sets, where the validation set is used to detect overfitting.

One problem with using ML for causal inference is that models typically do not produce consistent parameter estimates \citep{Mullainathan2017}. This is because ML models are designed for prediction accuracy, rather than inference. Individual parameter estimates are typically not objects of interest. Recently, however, the econometrics literature has started modifying ML methods for use within economics (\citealp{Varian2014}; \citealp{Mullainathan2017}; \citealp{Athey2017b}; \citealp{Athey2017c}; \citealp{Wager2017}). Notably, the debiased machine learning (DML) estimator, which was recently introduced by \cite{Chernozhukov17, Chernozhukov18}, allows for the unbiased estimation of a parameter of interest in the presence of a high dimensional and potentially nonlinear nuisance parameter.\footnote{The authors prove the root-N consistency of the DML estimator. They also show unbiasedness in finite samples using a Monte Carlo simulation. They also use the Monte Carlo exercise to demonstrate the necessity of both the orthogonalization and sample-splitting steps.} We will make use of DML to estimate the tax capitalization effect. Furthermore, we will couple the DML approach with a neural network architecture that was recently introduced in the machine learning literature by \cite{Cheng16}: the ``deep-wide'' network. The benefit of this choice of network architecture is that it allows for arbitrary nonlinear interactions between time-varying controls, such as public services and local characteristics, but uses a parsimonious specification for fixed effects that prohibits unintended interactions. In our case, the variation in the dependent variable is at the municipal-time level. We may want to include municipal and time fixed effects, but we do not want to allow them to interact, since there is no remaining variation in the data. Coupling DML with \cite{Cheng16} allows us to prevent such unintended interactions. The same cannot be achieved with standard deep neural network architectures, random forests, lasso regression, or elastic net regressions without deviating from the assumptions in \cite{Chernozhukov17, Chernozhukov18} or reducing computational efficiency.\footnote{One potential downside to using the architecture in \cite{Cheng16} is that it is unclear whether the convergence results for neural network estimators in \cite{FLM21} and \cite{Sch20} can be applied. This makes it difficult to assess whether the first stage convergence requirements of \cite{Chernozhukov17} are satisfied. For these reasons, we instead evaluate the finite sample properties of our estimator using Monte Carlo simulations that replicate the properties of our data.}

More formally, we treat the problem of estimating tax capitalization as a partially linear model, as described by \cite{Robinson88} and \cite{Chernozhukov17}. We follow the exposition and notation introduced in \cite{Chernozhukov17}. In Equation (\ref{eqn:partially_linear_1}), $p_{jt}$ is the square meter price of housing, $\tau_{jt}$ is the level of municipal taxes, $x_{jt}$ is a $k$-dimensional vector of confounders, $(x_{jt}^{1},...,x_{jt}^{k})$, and $u_{jt}$ is the disturbance term.

\begin{equation} \label{eqn:partially_linear_1}
p_{jt} = \tau_{jt} \theta_{0} + g_{0}(x_{jt}) + u_{jt}
\end{equation}

The functional form, $g_{0}(x_{jt})$, allows for the high-dimensional vector of confounders to have a direct and potentially nonlinear impact on $p_{jt}$. We assume that the disturbance term is zero in expectation, conditional on $x_{jt}$ and $\tau_{jt}$:

\begin{equation} \label{eqn:partially_linear_2}
E[u_{jt}|x_{jt},\tau_{jt}] = 0
\end{equation}

This setup also allows for the possibility that confounders may influence the determination of municipal tax rates:

\begin{equation} \label{eqn:partially_linear_3}
\tau_{jt} = m_{0}(x_{jt}) + v_{jt}
\end{equation}

We might expect, for example, that demographic variation across municipalities and time, captured in $x_{jt}$, will affect both $\tau_{jt}$ and $p_{jt}$. We allow for the confounders to do this through $m_{0}(x_{jt})$ and $g_{0}(x_{jt})$ separately.
Finally, we assume that the disturbance term is zero in expectation, conditional on the vector of confounders:

\begin{equation} \label{eqn:partially_linear_4}
E[v_{jt}|x_{jt}] = 0
\end{equation}

If $\tau_{jt}$ is conditionally exogenous, then $\theta_{0}$ may be interpreted as the unbiased and causal treatment effect.\footnote{While reverse causality is also possible, it is less likely in this setting, since the treatment is decided at the end of the previous period. Reverse causality would require politicians to increase taxes in anticipation of a future decline in house prices. Furthermore, this decline in house prices would have to be unforecastable.} In order to capture all plausible sources of confounding, $k$ must be large; however, $k$ being large is also problematic, as \cite{Chernozhukov17} show, because $k$ is typically assumed to grow slowly in the sample size. In our case, $k$ is large, which violates standard assumptions about nuisance parameter complexity.

\cite{Chernozhukov17} show that estimating the nuisance parameter, $\eta_{0} = (g_{0},m_{0})$, with machine learning techniques and then adding it to the estimating equations directly will lead to a bias in $\theta_{0}$. This arises from the use of regularization techniques, which allow machine learning methods to use a high number of covariates without overfitting. 
\cite{Chernozhukov17} demonstrate how to correct for this bias using orthogonalization and sample splitting. We start by splitting the dataset into two equal parts: $I$ and $I^{c}$. We use $I^{c}$ to estimate $\hat{m}_{0}$ and $\hat{g}_{0}$. We can then estimate the parameter of interest for a given split:

\begin{equation} \label{eqn: dml_equation}
\hat{\theta}_{0}(I^{c},I) = \left( \frac{1}{\bar{n}\bar{t}} \Sigma_{jt \in I} \hat{v}_{it} \tau_{jt} \right)^{-1} \frac{1}{\bar{n}\bar{t}} \Sigma_{jt \in J} \hat{v}_{jt} \biggl(p_{jt} - \hat{g}_{0}(x_{jt}) \biggr)
\end{equation}

Note that $\bar{n}$ and $\bar{t}$ are the number of municipalities and time periods in $I$. Furthermore, notice that $\hat{\theta}_{0}(I^{c},I)$ uses $\hat{m}_{0}$ and $\hat{g}_{0}$ estimated on the auxiliary sample, $I^{c}$, and $p_{jt}$ and $x_{jt}$ from the main sample, $I$. The final step swaps the main and auxiliary samples, computes $\hat{\theta}_{0}(I,I^{c})$, and calculates the mean:

\begin{equation}
\hat{\theta}_{0} = \frac{1}{2} [ \hat{\theta}_{0}(I^{c},I) + \hat{\theta}_{0}(I,I^{c})]
\end{equation}

Following \cite{Chernozhukov17}, we repeat this procedure 51 times and then compute the median parameter value. In the following subsection, we will discuss an alternative specification that allows for the use of fixed effects.

\subsection{Deep-Wide Network Architecture}

\begin{figure}[ht!]
  \caption{Deep-wide neural network architecture}
  \includegraphics[width=\linewidth]{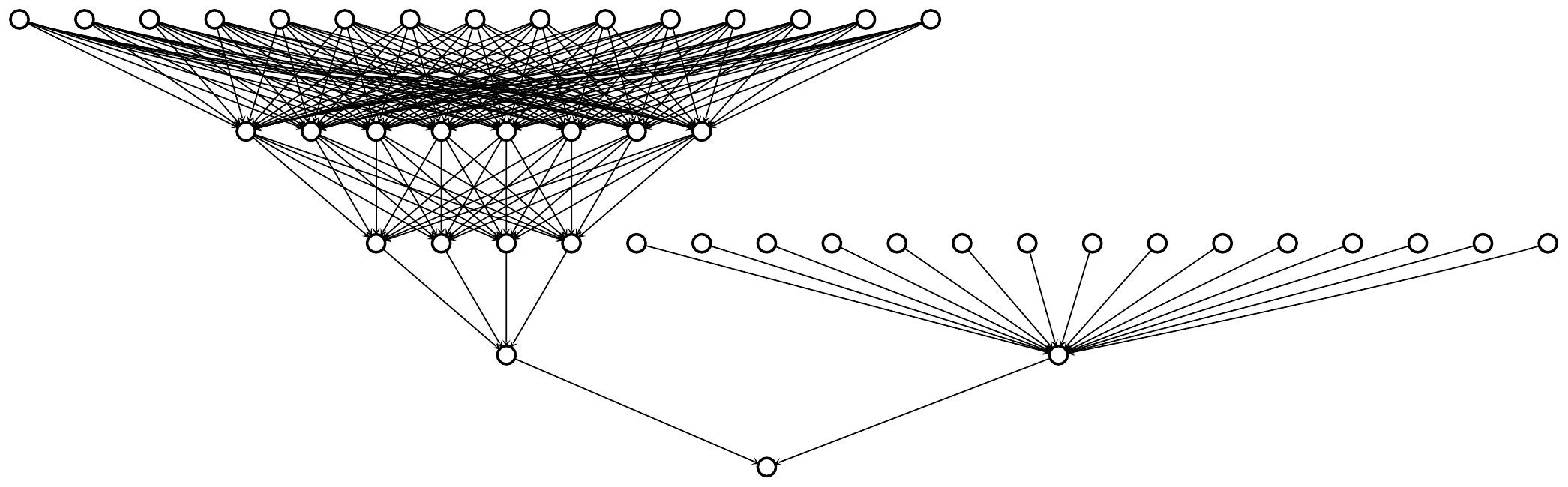}
  \label{fig:deep_wide}
\footnotesize
$Notes$: The figure above shows a deep-wide neural network. The ``deep'' part of the network is shown on the left and the ``wide'' part is shown on the right. Note that the deep component allows for arbitrary nonlinear interactions between control variables, which are captured by activation functions that are applied at each node. The wide part of the network efficiently parameterizes the fixed effects by embedding them in a linear model. The deep and wide components of the model are then trained jointly using the adaptive moment optimizer.
\end{figure}

\cite{Chernozhukov17} point out that DML can be performed with a wide variety of machine learning estimators, including random forests, penalized linear regression, and neural networks. We make use of a neural network to construct the DML estimator in this paper, but we use an atypical network architecture that is ideally suited to our problem and a large class of estimation problems in economics. We do this by importing the ``deep-wide'' network that was recently introduced in the machine learning literature by \cite{Cheng16}. Its architecture consists of two subnetworks: 1) a deep neural network; and 2) a ``wide'' or linear subnetwork. Figure \ref{fig:deep_wide} depicts the network's architecture. The deep subnetwork is shown on the left and the wide subnetwork is shown on the right.

In our implementation, the high dimensional vector of municipal characteristics and public services are used as inputs to the deep subnetwork, allowing for arbitrary nonlinear interactions between the control variables.\footnote{Our problem differs from the one in \citet{Cheng16}, which is why we adopt their deep-wide network structure, but do not replicate the specific architecture they adopt in the deep subnetwork.} This would not be possible in a penalized linear regression model, such as a lasso, which does not permit nonlinearities. Furthermore, while other machine learning methods, such as random forests and standard neural network architectures do allow for nonlinearities, they also permit unintended interactions between fixed effects. In a deep-wide network, the fixed effects are handled as inputs to the linear subnetwork, allowing for an efficient parameterization and also preventing unintended interactions.\footnote{Since we cannot partition the sample by both municipality and time, the estimated time effects in each split will depend on the set of associated municipalities drawn for each time period. Similarly, the municipality fixed effects will depend on the times for which they are drawn. To account for this, we repeat the estimation procedure 51 times and then select the median result, as is done in \cite{Chernozhukov17}. We also use Monte Carlo simulations to explore the finite sample properties of our estimator and find that this is unlikely to have an impact on our results. See \cite{CKMS21} for an alternative approach and \cite{KT19} for convergence results in dynamic panel settings where the estimator has properties similar to \cite{Chernozhukov17}.} Finally, our choice of architecture also allows for joint estimation of the deep and wide subnetworks, and can be adapted to allow for nonlinear dependence between the deep and wide network outputs.\footnote{We use a linear activation in the final layer of the network. A simple modification of the network that uses a nonlinear activation in the layer in which the deep and wide subnetworks are merged would allow for nonlinear dependence.} This provides an advantage over using an ensemble of separately-estimated fixed effect and deep neural network models or residualizing house prices with respect to fixed effects.\footnote{At a minimum, the approach we adopt provides improvements in computational efficiency. It also allows for increased flexibility in model specification, including through the nonlinear dependence of subnetwork outputs, which cannot be achieved through residualization. Our approach also has the advantage of operating within the framework of DML and has all of the properties demonstrated in \cite{Chernozhukov17, Chernozhukov18}.}

To evaluate the empirical framework we adopt in this paper, we design a Monte Carlo experiment to compare DML with a deep-wide network (DML-DW) to alternative empirical strategies. In particular, we specify data generating processes for both the dependent variable (house prices) and the variable of interest (municipal taxes) that consist of time fixed effects, municipal fixed effects, and a nonlinear function of local variables with region and time variation, which corresponds to the nuisance parameter. The processes are consistent with the aforementioned partially linear model setup and are described in full detail in Section \ref{monte_carlo} of the Appendix.

\begin{figure}[ht!]
  \caption{Kernel density plots of coefficient estimates}
  \includegraphics[width=\linewidth]{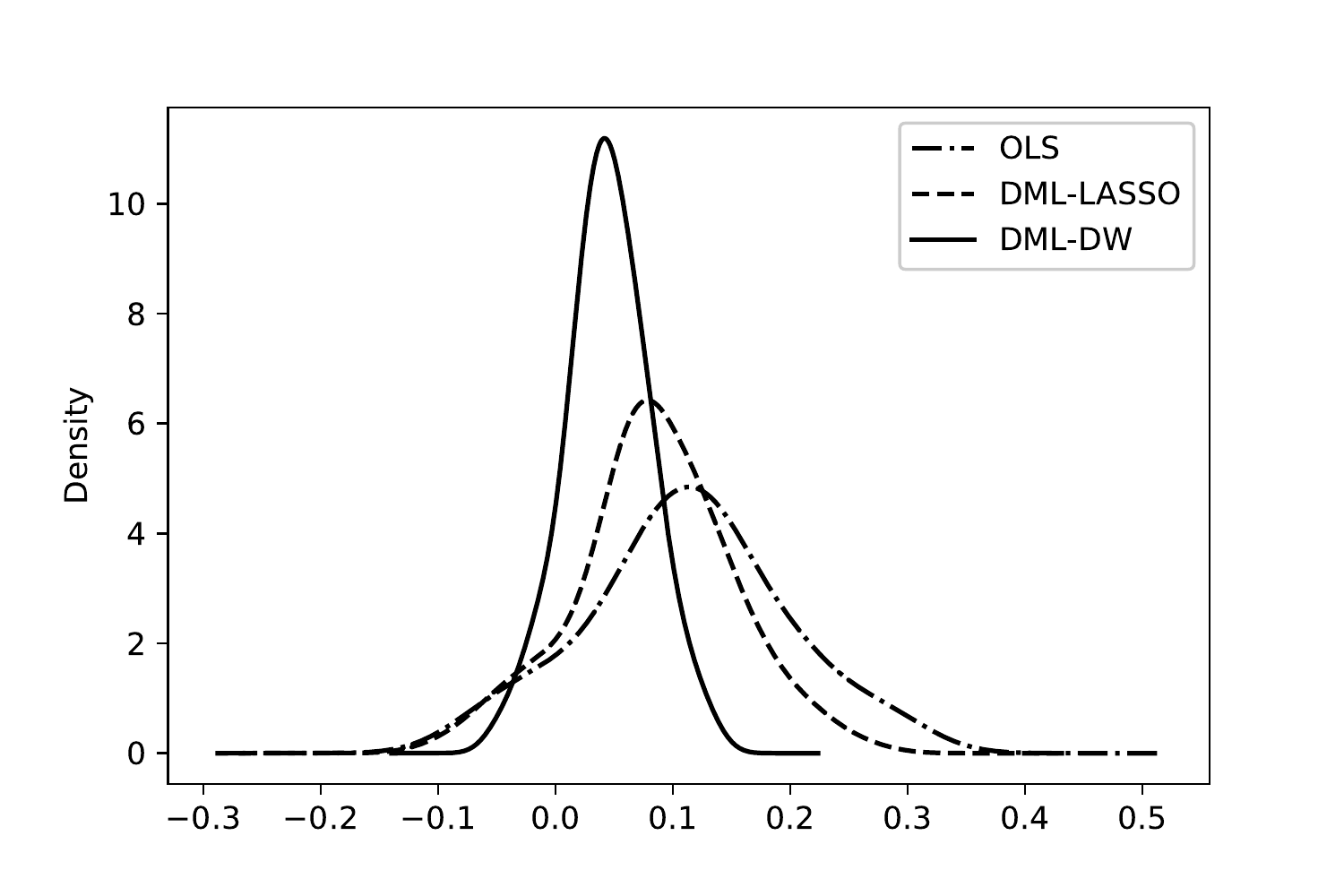}
  \label{fig:monte_carlo}
\footnotesize
$Notes$: The figure above shows kernel density plots of estimated tax capitalization coefficient biases from a Monte Carlo simulation. In each case, we randomly generate data and then use it to estimate the parameter of interest using OLS, DML with a deep-wide network, and DML with a LASSO. Note that double ML with a deep-wide network is approximately unbiased; whereas, OLS and DML-LASSO are both biased and have a higher variance of estimates. See Section \ref{monte_carlo} for the details of the exercise.
\end{figure}

We then compare the bias in DML-DW estimates to two alternative empirical strategies: 1) OLS with fixed effects and a random subset of the controls; and 2) DML with a LASSO regression. Alternative (1) is intended to replicate much of the empirical literature, which uses OLS and only a random subset of local controls. Alternative (2) makes use of DML, the full set of fixed effects and controls, and an estimation strategy that employs regularization, but does not permit nonlinearities in the controls. As shown in Figure \ref{fig:monte_carlo}, DML-DW outperforms both alternatives in the simulation environment we consider, which is constructed to replicate the empirical features of the tax capitalization problem. For the full details of the Monte Carlo experiments, see Section \ref{monte_carlo} of the Appendix.

We implement a custom version of the DML-DW algorithm in TensorFlow. Our approach allows for flexibility in the choice of network architecture, including the number of hidden layers, the number of nodes in each layer, the type of activation functions, the type of regularization employed,\footnote{We use L2 regularization in our empirical exercises; however, our code also allows for the use of L1 regularization and dropout.} and the optimization algorithm employed. We also allow for automatic selection of the number of training epochs\footnote{The training process in machine learning typically divides the sample into batches, which are fed into the optimizer in sequence. Training a model entails stepping over the complete sample tens or hundreds of times. Each pass over the entire sample is referred to as an epoch.} based on out-of-sample prediction performance to prevent overfitting. The technique could be employed to estimate parameters of interest in a large class of problems with a high number of covariates and fixed effects.
\section{Results}\label{Res}

In this section, we discuss the empirical results, starting with our findings for tax capitalization. We then move on to migration, which was emphasized by \cite{Tiebout56} as the mechanism through which tax changes capitalize into house prices. We then discuss how public services and crime affect house prices. Finally, for each empirical exercise, we provide separate estimates for urban areas only. Within these areas, municipal competition is higher, which should lead to increased tax capitalization according to \cite{Tiebout56}.

\subsection{Tax Capitalization}\label{Rex_tax}

We first examine tax capitalization in a simple empirical setting with OLS. Our baseline specification, shown in Equation (\ref{eqn:OLS_1}), regresses the square meter price of housing at the municipal level, $p_{jt}$, on the municipal tax rate, $\tau_{jt}$; municipal fixed effects, $\gamma_{j}$; and yearly time fixed effects, $\eta_{t}$. Notice that we do not use time-varying controls in this specification.\footnote{Given the Monte Carlo simulation results, we concentrate on DML-DW as our preferred DML specification, but also include OLS specifications as a benchmark, since they are commonly used in the literature.}

\begin{equation} \label{eqn:OLS_1}
p_{jt} = \tau_{jt} \theta_{0} + \gamma_{j} + \eta_{t} + u_{jt}
\end{equation}

We also use a separate specification that includes a set of time-varying controls, $x_{jt}$, that is standard in the literature. This includes spending per pupil, a measure of grades, and the number of violent crimes per 100,000 inhabitants. This specification is shown in Equation (\ref{eqn:OLS_2}). Throughout this section, we use standardized versions of the square meter house price, the municipal tax rate, and time-varying controls. All results should be interpreted as the change in the dependent variable in standard deviations associated with an increase in the variable of interest in standard deviations.

\begin{equation} \label{eqn:OLS_2}
p_{jt} = \tau_{jt} \theta_{0} + \gamma_{j} + \eta_{t} + \beta x_{jt} + u_{jt}
\end{equation}

Table \ref{tab:tax_elasticity_regressions} shows our baseline set of estimates for tax capitalization. Column (1) uses the specification in Equation (\ref{eqn:OLS_1}) and column (2) uses the specification in Equation (\ref{eqn:OLS_2}). Both columns report standard errors that are clustered at the municipal level.\footnote{We are unaware of any algorithm that allows for clustering errors in the DML-DW specification. The most closely related work that provides a clustering algorithm is \cite{Chi18}.} Note that we cannot use a specification with municipality-year fixed effects, since the variation in the dependent variable is at the municipality-year level. The only confounders omitted are the time-varying public goods and services, and time-varying municipal characteristics that are novel to this paper. 

\begin{table}[ht!]
\caption{Impact of municipal income tax level on house prices}
\begin{center}
\resizebox{0.70\linewidth}{!}{
\begin{tabular}{lcccc}
\hline
& (1) & (2) & (3)  \\
& (OLS) & (OLS) & (DML-DW) \\
\hline
\hline
$\mbox{municipal\_tax}_{it}$  & -0.1145** & -0.1236*** & -0.2626*** \\
 & (0.051) & (0.051)  & (0.031) \\
\hline
Standard Controls & \xmark & \Checkmark & \Checkmark \\
Housing & \xmark & \xmark & \Checkmark \\
Migration & \xmark & \xmark & \Checkmark \\
Political & \xmark & \xmark & \Checkmark \\ 
Labor & \xmark & \xmark & \Checkmark \\
Demographic & \xmark & \xmark & \Checkmark \\
Economic & \xmark & \xmark & \Checkmark \\
Public Finance & \xmark & \xmark & \Checkmark \\
Public Service Inputs & \xmark & \xmark & \Checkmark \\
Public Service Ouputs & \xmark & \xmark & \Checkmark \\
Schooling & \xmark & \xmark & \Checkmark \\
\hline
Year FE & YES & YES & YES \\
Municipal FE & YES & YES & YES \\
Standard Errors & CL & CL & - \\
\hline
Adj. R-squared   & 0.9531 & 0.9535  & - \\
N & 1764 & 1764 & 1764 \\
\hline
\end{tabular}}
\end{center}
\footnotesize
$Notes$: The dependent variable is the standardized square meter price of a villa in municipality (kommun) $i$ and year $t$. The regressor of interest is the standardized municipal-level income tax rate, which is observed annually. Standard controls include measurements of spending per pupil, education, and violent crime. Fixed effects are applied at the municipal level. DML refers to double-debiased machine learning. DW indicates that a deep-wide network was used. Note that a \Checkmark indicates that the referenced group of control variables was included. CL indicates that standard errors are clustered at the municipal level. * $p<.1$, ** $p<.05$, *** $p<.01$.
\label{tab:tax_elasticity_regressions}
\end{table}

Next, we employ the DML algorithm to produce estimates that incorporate both the full set of fixed effects and also the time-varying controls. We follow the approach outlined in Section \ref{Est}. The results given in column (3) use DML, coupled with the deep-wide neural network architecture we import from the machine learning literature (DML-DW). Notice that the checkmarks (\Checkmark) indicate whether a group of time-varying controls is included. Column (3) incorporates the full set of control group categories, which includes housing, migration, politics, labor, demographics, economics, public finance, public service inputs, and public service outputs. This also means that DML-DW removes the variation in taxes attributable to those same controls, including changes in local political preferences, economic conditions, or budgetary considerations. No other work in the literature is able to achieve identification in a similar manner, but must instead rely on the assumption that a given policy change was exogenous.

We modify the setup described in Section \ref{Est_DML} by incorporating the municipality and year fixed effects into the nuisance parameter, as shown in Equation (\ref{eqn:modified_partially_linear_1}). The original process is given by Equation (\ref{eqn:partially_linear_1}). Furthermore, we still allow the time-varying controls to have a potentially nonlinear impact on the square meter price through $\zeta_{g}$, while restricting the fixed effects to enter linearly into the model through $\gamma_{j}$ and $\eta_{t}$.\footnote{We also perform DML with the same specification, but using different ML methods, including lasso regression, elastic net regression, random forests, and standard neural network architectures. Lasso regression provides results that are most similar to our main specification with the deep-wide network architecture; however, the results are highly sensitive to the choice of regularization parameter. Similar to \cite{Chernozhukov17, Chernozhukov18}, we also find that the standard errors for DML with a lasso regression are higher than for other ML methods.} Our approach allows for the possibility that $g_{0}$ is a nonlinear combination of the fixed effects and $\zeta_{g}$; however, in the applications we consider, we will assume that it is not.

\begin{equation} \label{eqn:modified_partially_linear_1}
p_{jt} = \tau_{jt} \theta_{0} + g_{0}(\zeta_{g}(x_{jt}) + \gamma_{j} + \eta_{t}) + u_{jt}
\end{equation}

Similarly, we modify the estimating equation for $\tau_{jt}$, given in Equation (\ref{eqn:partially_linear_3}) by incorporating municipality and time fixed effects, as is shown in Equation (\ref{eqn:modified_partially_linear_2}).

\begin{equation} \label{eqn:modified_partially_linear_2}
\tau_{jt} = m_{0}(\zeta_{m}(x_{jt}) + \delta_{j} + \xi_{t}) + v_{jt}
\end{equation}

The remainder of the DML algorithm is executed as described in Section \ref{Est}. The ``deep'' side of the network contains two hidden layers, one with 16 nodes and one with 8 nodes. We use rectified linear unit activation functions for all hidden layers and a linear activation in the output layer. The architecture for the deep side is also designed to prevent overfitting. We accomplish this in two ways. First, we apply regularization to each hidden layer of the network.\footnote{We use L2 regularization with a 0.05 penalty, which penalizes the squared values of the weights connecting the hidden layers.} Second, we divide the dataset into training (80\%) and validation (20\%) splits, and select the number of epochs at which the train and validation samples have approximately the same loss function value. We use a mean absolute error loss function and the adaptive moment optimizer.\footnote{The adaptive moment estimator or ``adam'' was introduced by \cite{Kingma2014}. It is one of the most commonly-used optimization algorithms in machine learning and has several attractive properties for the class of problems we try to solve. First, it applies different step sizes to each component of the gradient, which is useful for high-dimensional optimization problems. Second, it has a parsimonious set of hyperparameters, which can easily be interpreted and tuned. Third, it has good convergence properties. And fourth, it has been demonstrated to perform well empirically in a large variety of applications. Using an MAE loss results in a more stable training process, but is not important for our results. We use standard training parameters for adam: a learning rate of 0.001, a first moment exponential decay rate of 0.9, and a second-moment exponential decay rate of 0.999.} Furthermore, we apply the DML step 51 times and select the median estimate, similar to what is done in \cite{Chernozhukov17, Chernozhukov18}.

First, notice that our baseline OLS estimate of tax capitalization is negative, which is consistent with the previous literature. Column (1) indicates that a one standard deviation increase in the level of the municipal tax is associated with a 0.1145 standard deviation decrease in the square meter house price. In column (2), the inclusion of standard controls from the literature increases the magnitude of the estimate from -0.1145 to -0.1236 and increases its significance from the 5\% level to the 1\% level. Finally, our baseline DML-DW estimate, shown in column (3), yields a tax capitalization impact that is roughly twice as large as our estimate using OLS with fixed effects. In particular, we find that a one standard deviation increase in the income tax at the municipal level is associated with a 0.26 standard deviation decrease in the square meter price of housing. This is equivalent to a square meter price reduction of 2900 SEK (or 446 USD at 2013 exchange rates).

Beyond our baseline estimate, we can also see that selective inclusion of control groups results in range of coefficient estimates from -0.194 to -0.34. This is shown in Table \ref{tab:tax_elasticity_regressions_control_groups}, which provides the tax capitalization estimates using just one of the control groups in each of the columns. Furthermore, notice that we divide public services into two control groups: inputs and outputs. \cite{Oates1969} was the first to point out that using public expenditures (inputs), as \cite{Tiebout56} suggested, would only imperfectly capture outputs, which was the true object of interest. While \cite{Oates1969} did not use measures of public service and good outputs due to data availability, later work in the literature (\citealp{Rosen77}; \citealp{Hanushek86}; and \citealp{Hanushek96}) confirmed the importance of outputs. We find that the inclusion of outputs leads to a larger increase in the magnitude of the impact of tax capitalization, which supports the claim that inputs are an imperfect proxy for outputs. We will revisit this question again when we measure the impact of public goods and services.

Furthermore, the selective inclusion of control groups allows us to identify which confounders are likely to have the largest effect on the bias in tax capitalization estimates. In particular, we find that housing variables (including measures of supply), public service outputs, and public finance variables have the largest impact on tax capitalization estimates. The importance of housing market and public finance variables suggests that the joint determination of taxes and house prices is important for the empirical setting we consider and could plausibly be a driver of bias in existing estimates in the literature. Even in quasi-experimental settings and in work that exploits border discontinuities, the exclusion of public finance and housing market variables could be problematic, since they could be partially responsible for the shifts in local tax rates.

\begin{sidewaystable}
\caption{Impact of control groups on tax capitalization estimates}
\resizebox{\linewidth}{!}{
\centering
\begin{tabular}{lcccccccccc}
\hline
& (1) & (2) & (3) & (4) & (5) & (6) & (7) & (8) & (9) & (10) \\
& (DML-DW) & (DML-DW) &  (DML-DW) &  (DML-DW) &  (DML-DW) &  (DML-DW) &  (DML-DW) &  (DML-DW) &  (DML-DW) &  (DML-DW) \\
\hline
\hline
$\mbox{municipal\_tax}_{it}$ & -0.3438*** & -0.1942*** & -0.2173*** & -0.1933*** & -0.2092*** & -0.2176*** & -0.2982*** & -0.2587*** & -0.3173*** & -0.2184***\\
 & (0.0483) & (0.0311) & (0.0280) & (0.0348) & (0.0374)  & (0.0336) & (0.0353) & (0.0335) & (0.0393) & (0.0323) \\
\hline
Housing & \Checkmark & \xmark & \xmark & \xmark & \xmark & \xmark & \xmark & \xmark & \xmark & \xmark \\
Migration & \xmark & \Checkmark & \xmark & \xmark & \xmark & \xmark & \xmark & \xmark & \xmark & \xmark \\
Political & \xmark & \xmark & \Checkmark & \xmark & \xmark & \xmark & \xmark & \xmark & \xmark & \xmark \\ 
Labor & \xmark & \xmark & \xmark & \Checkmark & \xmark  & \xmark & \xmark & \xmark & \xmark & \xmark \\
Demographic & \xmark & \xmark & \xmark & \xmark & \Checkmark & \xmark & \xmark & \xmark & \xmark & \xmark \\
Economic & \xmark & \xmark & \xmark & \xmark & \xmark & \Checkmark & \xmark & \xmark & \xmark & \xmark \\
Public Finance & \xmark & \xmark & \xmark & \xmark & \xmark & \xmark & \Checkmark & \xmark & \xmark & \xmark \\
Public Service Inputs & \xmark & \xmark & \xmark & \xmark & \xmark & \xmark & \xmark & \Checkmark & \xmark & \xmark \\
Public Service Ouputs & \xmark & \xmark & \xmark & \xmark & \xmark & \xmark & \xmark & \xmark & \Checkmark & \xmark \\
Schooling & \xmark & \xmark & \xmark & \xmark & \xmark & \xmark & \xmark & \xmark & \xmark & \Checkmark \\
\hline
Year FE & YES & YES & YES & YES & YES & YES & YES & YES & YES & YES \\
Municipal FE & YES & YES & YES & YES & YES  & YES & YES & YES & YES & YES \\
\hline
N & 1764  & 1764  & 1764   & 1764  & 1764 & 1764  & 1764  & 1764   & 1764  & 1764 \\
\hline
\end{tabular}}
\scriptsize
$Notes$: The dependent variable is the standardized square meter price of a villa in municipality (kommun) $i$ and year $t$. The regressor of interest is the standardized municipal-level income tax rate, which is observed annually. Fixed effects are applied at the municipal level. DML refers to double-debiased machine learning. DW indicates that a deep-wide network was used. Note that a \Checkmark indicates that the referenced group of control variables was included. * $p<.1$, ** $p<.05$, *** $p<.01$.
\label{tab:tax_elasticity_regressions_control_groups}
\end{sidewaystable}

We next split our sample along urban-rural lines. The urban subsample contains all observations associated with municipalities in Sweden's most populous counties: Stockholm, Sk{\aa}ne, and V\"astra G\"otaland. These counties are also more population-dense and municipality-dense than the remaining 18. Furthermore, all variables are re-standardized within the urban subsample. Table \ref{tab:tax_elasticity_population_density} provides results for the DML-DW estimator with municipality and time fixed effects, as well as a complete set of time-varying controls. Note that the estimated impact for rural areas is approximately zero; whereas the impact for urban areas is roughly four times greater in magnitude than the baseline estimate for the full sample. This accords well with \cite{Tiebout56}, which argues that the impact should be greatest where moving costs are low. In this case, the cost of moving between municipalities should be lowest in dense urban areas, where many alternatives are available and where moving will not typically require a change in employment.

\begin{table}[ht!]
\caption{Impact of municipal income tax level on house prices by population density}
\begin{center}
\resizebox{0.85\linewidth}{!}{
\footnotesize
\begin{tabular}{lcc}
\hline
& (1) & (2)  \\
& (DML-DW) & (DML-DW) \\
\hline
\hline
$\mbox{municipal\_tax}_{it}$  & -1.043*** & -0.008*** \\
 & (0.0216)  & (0.003)  \\
\hline
Year FE & YES & YES \\
Municipal FE & YES & YES \\
Time-Varying Controls & YES & YES \\
Counties & Urban & Rural \\
\hline
N & 658  & 1106  \\
\hline
\end{tabular}}
\end{center}
\footnotesize
$Notes$: The dependent variable is the standardized square meter price of a villa in municipality (kommun) $i$ and year $t$. The regressor of interest is the standardized municipal-level income tax rate, which is observed annually. We use two subsamples: 1) Urban and 2) Rural. DML refers to double-debiased machine learning. DW indicates that a deep-wide network was used. * $p<.1$, ** $p<.05$, *** $p<.01$.
\label{tab:tax_elasticity_population_density}
\end{table}

\subsection{Migration}

Following \cite{Cebula78} and \cite{Banzhaf08}, we further examine the claim in \cite{Tiebout56} that households will ``vote with their feet.'' We do this by directly measuring the impact of municipal taxes on net migration into the municipality. A positive rate of net migration indicates that more individuals are entering the municipality than are leaving it in a given year. Our results are given in Table \ref{tab:net_migration}. In column (1), we use DML-DW with time and municipality fixed effects, as well as the full set of time-varying controls. This yields an impact of -0.062, which is significant at the 1\% level. When we limit ourselves to urban areas only in column (2), this estimate nearly doubles to -0.1063, suggesting that the underlying mechanism in \cite{Tiebout56} for tax capitalization may, indeed, be supported empirically. Furthermore, it is plausible that the long-run effect could be larger, since many households might be unable to move within a year of the tax change's announcement.

\begin{table}[ht!]
\caption{Impact of taxes on net migration}
\begin{center}
\resizebox{0.85\linewidth}{!}{
\begin{tabular}{lcccc}
\hline
& (1) & (2) \\
& (DML-DW) & (DML-DW) \\
\hline
\hline
$\mbox{municpal\_tax}_{it}$  & -0.062*** & -0.1063***  \\
& (0.005) & (0.009) \\
\hline
Year FE & YES & YES \\
Municipal FE & YES & YES \\
Standard Controls & YES & YES \\
Time-Varying Controls & YES & YES \\
Counties & ALL &  URBAN \\
\hline
N & 1764 & 658 \\
\hline
\end{tabular}}
\end{center}
\small
$Notes$: The dependent variable is standardized net migration in (kommun) $i$ and year $t$. The regressor of interest is the standardized spending per pupil at the municipality-level, which is observed annually. DML refers to double-debiased machine learning. DW indicates that a deep-wide network was used. * $p<.1$, ** $p<.05$, *** $p<.01$.
\label{tab:net_migration}
\end{table}

\subsection{Education}

\begin{table}[hb!]
\caption{Impact of educational inputs and outputs on house prices}
\resizebox{\linewidth}{!}{%
\begin{tabular}{lcccc}
\hline
& (1) & (2) & (3) & (4) \\
&  (DML-DW) & (DML-DW) &  (DML-DW) & (DML-DW) \\
\hline
\hline
$\mbox{spending\_per\_pupil}_{it}$  & -0.0297*** & -0.0248*** & -  & - \\
& (0.008)  & (0.006) & -  & - \\
$\mbox{grades}_{it}$  & - & - & 0.0011 &  0.0211*** \\
&  - & - & (0.005) & (0.009) \\
\hline
Year FE & YES & YES & YES & YES \\
Municipal FE & YES & YES & YES & YES \\
Time-Varying Controls & YES & YES & YES & YES \\
Counties & ALL & URBAN & ALL & URBAN \\
\hline
N & 1764 & 658 & 1764 & 658 \\
\hline
\end{tabular}}
\small
$Notes$: The dependent variable is the standardized square meter price of a villa in municipality (kommun) $i$ and year $t$. The regressor of interest is the standardized spending per pupil at the municipality-level and a measure of grades at the municipal-level, both of which are observed annually. DML refers to double-debiased machine learning. DW indicates that a deep-wide network was used. * $p<.1$, ** $p<.05$, *** $p<.01$.
\label{tab:education}
\end{table}

We next make an attempt to directly measure the impact of public services on house prices. We start by examining education, since this is one of the most frequently tested public services in the literature (see, e.g., \citealp{Haurin96}; \citealp{Black99}; \citealp{Downes2002}; \citealp{Barrow04}; \citealp{Cheshire04}; \citealp{Bayer07}; \citealp{Ries2010}). In most work, either spending per pupil is used an input or grades are used as an output. Table \ref{tab:education} shows our results for spending per pupil and grades.

Since spending per pupil is an input, we might expect that the positive effect found in the literature arises from the omission of variables for schooling output, such as grades and quality measures, which comove with spending. In columns (1) and (2), we make use of our extensive set of time-varying municipal controls to estimate the impact of spending per pupil in isolation. In particular, in column (1), we use DML-DW and include year and municipal fixed effects, as well as time-varying controls. This includes educational outputs, such as grades. We find that a one standard deviation increase in spending is associated with a 0.0297 standard deviation decrease in the square meter price of housing. This is what we might expect, given that we are able to include controls for educational outputs and other highly correlated public services.

We next measure the impact of outputs, rather than inputs, using grades as a measure. The results are shown in Table \ref{tab:education}. Column (3) uses DML-DW, coupled with year and municipal fixed effects, and time-varying controls, but yields no significant effect. If we again restrict ourselves to urban areas, the size of the effect rises to 0.021 and becomes significant at the 1\% level. This effect, however, remains quite small, even in urban counties, suggesting that much of the apparent positive association between educational outcomes and house prices might actually be capturing comovement with omitted municipal characteristics. It is also plausible, of course, that the long-run impact of improvements in grades could be substantially higher.

\subsection{Crime}

In addition to measuring tax and public service capitalization, we also examine the impact of crime on house prices in a final exercise. While this question differs slightly from the core aim of this paper, it is closely related and allows us to contribute to a large literature on the subject (see, e.g., \citealp{Thaler78}; \citealp{Reinhard81}; \citealp{Blomquist88}; \citealp{Haurin96}; \citealp{Gibbons04}; \citealp{Linden08}) using our novel dataset, coupled with the DML-DW approach. Our estimates are given in Table \ref{tab:crime}. The dependent variable is the standardized square meter price of housing and the variable of interest is the standardized number of violent crimes per 100,000 inhabitants. Column (1) performs DML-DW with a full set of time-varying controls, and year and municipal fixed effects. We find that a one standard deviation increase in crime lowers house prices by -0.0154 standard deviations. Limiting the sample to urban areas increases the magnitude of the effect to -0.054, providing further confirmation of the Tiebout hypothesis. All estimates are significant at the 1\% level. Additionally, all variables are re-standardized in the urban subsample. Overall, our findings suggest that the impact of violent crime on property prices is negative, but small in the short run. As with educational outputs, it is plausible that the impact of crime could be more substantial in the long-run.

\begin{table}[ht!]
\caption{Impact of crime on house prices}
\begin{center}
\resizebox{0.65\linewidth}{!}{%
\begin{tabular}{lcc}
\hline
& (1) & (2) \\
& (DML-DW) & (DML-DW) \\
\hline
\hline
$\mbox{crime}_{it}$ &  -0.0154*** & -0.0540*** \\
 & (0.003) & (0.008) \\
\hline
Year FE  & YES & YES \\
Municipal FE & YES & YES \\
Standard Controls & YES & YES \\
Time-Varying Controls  & YES & YES \\
Counties & ALL & URBAN \\
\hline
N & 1764 & 658   \\
\hline
\end{tabular}}
\end{center}
\small
$Notes$: The dependent variable is the standardized square meter price of a villa in municipality (kommun) $i$ and year $t$. The regressor of interest is the standardized number of violent crimes per 100,000 inhabitants. DML refers to double-debiased machine learning. DW indicates that a deep-wide network was used. * $p<.1$, ** $p<.05$, *** $p<.01$.
\label{tab:crime}
\end{table}
\section{Conclusion}\label{Con}

\citet{Tiebout56} first argued that the free-rider problem for local public goods could be resolved entirely through preference revelation. That is, households could ``vote with their feet'' by moving to a community that offered their preferred bundle of tax rates and public goods. \cite{Oates1969} followed \citet{Tiebout56} by providing a first empirical test of the theory by measuring the impact of local taxation and expenditures on housing values. He found that approximately two thirds of changes in property taxation were capitalized into prices. In addition to this, his work spawned a large empirical literature on the capitalization of local taxes into housing values.

Since \cite{Oates1969}, the empirical literature has struggled to obtain unbiased estimates of tax capitalization. The main problem is that measures of public services are typically not available at the local level. Consequently, the omission of public service controls has likely lead to a substantial bias in estimates, as mentioned in the literature and documented in \citet{Wales74} and \citet{Palmon1998}. Furthermore, even beyond the omission of public service controls, the literature lacks controls for variables that jointly determine tax policy and house prices, including housing market variables and public finance variables. 

We contribute to the literature in three ways. First, we assemble a novel and exhaustive dataset of house prices, local public goods and services, local characteristics, and local taxes for Sweden. In total, we have 947 time-varying local controls. Our dataset spans the full set of 290 municipalities and the period between 2010 and 2016. In addition to the time-varying controls, the high degree of time and geographic variation allows us to include a large number of fixed effects to sweep out confounding variation. Second, we use the recently introduced double machine learning estimator from \cite{Chernozhukov17, Chernozhukov18}, which enables the estimation of a parameter of interest in the presence of a potentially nonlinear and high dimensional nuisance parameter. We also couple the method with a neural network architecture that was introduced in \cite{Cheng16}, called a ``deep-wide'' network. The combination of DML and a deep-wide network architecture enables us to estimate tax capitalization in a specification where we allow for nonlinear dependence on controls, but restrict dependence on fixed effects to be linear. This refinement is likely to be generally useful for estimation problems that involve a high number of controls and fixed effects. And third, we use our novel dataset and econometric framework to estimate the impact of taxes, public service inputs, public service outputs, and crime on house prices. We also test the impact of taxes on migration, which provides insight into the mechanism underlying the Tiebout hypothesis. 

Overall, we find that excluding public service, public finance, and housing market controls leads to a substantial downward bias the magnitude of tax capitalization estimates. The inclusion of time-varying local controls, coupled with econometric methods that are capable of handling a high vector of covariates, yields a doubling of the estimated reduction in house prices in response to an increase in taxes. We also show that this effect is four times as large in urban areas, where municipal competition is highest, as \cite{Tiebout56} suggests. In addition to this, we test the underlying mechanism in \citet{Tiebout56} more directly by estimating the impact of municipal taxes on migration. We find a small effect using the entire sample, but it nearly doubles when we exclusively use the subsample of urban municipalities. We also show that public service inputs, such as spending per pupil, actually have a negative effect when outputs are properly controlled for; whereas the impact of outputs remains positive. This builds upon the sub-literature that emphasizes the importance of public service outputs (see \citealp{Oates1969}; \citealp{Rosen77}, \citealp{Hanushek86}; and \citealp{Hanushek96}). Finally, we show an application of our methodology to the estimation of crime effects on house prices, following vast literature on that topic (see, e.g., \citealp{Thaler78}; \citealp{Reinhard81}; \citealp{Blomquist88}; \citealp{Haurin96}; and \citealp{Gibbons04}; \citealp{Linden08}.) 

The issue of tax and public service capitalization into housing values has attracted a lot of attention in the literature since Tiebout's seminal paper in 1956. On a more practical note, this literature also has important implications for public policy at the local level. Knowing what drives migration and how specific public services are valued by households may enable local governments to better meet the needs of their constituents. Furthermore, from an individual's perspective, it may be useful to know how house prices respond to changes in local taxation. This knowledge may be particularly valuable in regions where there is a high level of municipal competition, where individuals can choose from many housing locations with similar commute lengths. While past work on tax and public service capitalization suffered from either a lack or a restricted set of public service and local characteristic controls, advances in data collection and availability will likely alleviate this problem for many countries in the future. In this paper, we propose an alternative framework for obtaining causal and unbiased tax capitalization estimates that will have increasing value in the future, as both the cultivation of ``big data'' sources and the use of machine learning become more commonplace in economics.

\newpage

\bibliographystyle{ecta}
\bibliography{housingbibtex}

\newpage
\appendix
\section{Appendix}\label{app}

\subsection{Control Series Subcategory List}\label{app_series}

This section provides a list of all categories of controls, along with all of their respective subcategories. ``Demographics,'' for instance, is a category of control and ``births and deaths'' is a subcategory that contains multiple series. Note that we do not use all categories of controls in the empirical exercises. Additionally, some control groups are subsets or combinations of the groups listed below. The control group ``schooling,'' for instance, corresponds to the subset of the ``educational'' controls that are specifically related to schools, teachers, and students.

\begin{enumerate}
\itemsep 0em

\small

\item \underline{\textbf{Demographics.}}

\begin{itemize}
\itemsep 0em
\item Births and deaths.
\item Expenses due to population change.
\item Foreign-born residents.
\item Life expectancy.
\item Parental education.
\item Personal assistance and disability.
\item Residents by age (number and share).
\item Resident education level by grade.
\item Student demographics.
\item Total inhabitants.
\item Women (share).
\end{itemize}

\vspace{2mm}

\item \underline{\textbf{Disability \& Elder Care.}}

\begin{itemize}
\itemsep 0em
\item Elder care compensation.
\item Elder care and disability costs.
\item Elder care and disability employment.
\item Elder care and disability revenue.
\item Elder dependence on care.
\end{itemize}

\vspace{2mm}

\item \underline{\textbf{Economic.}}

\begin{itemize}
\itemsep 0em
\item Company registrations per 1000 persons.
\item Median salary.
\end{itemize}

\vspace{2mm}

\item \underline{\textbf{Education.}}

\begin{itemize}
\itemsep 0em
\item Adult education costs.
\item Cost of education by expense.
\item Cost of education by grade.
\item Cost of education by program.
\item Cost and income from preschool.
\item Educational revenue and income.
\item Gender parity in education.
\item Immigrant education costs.
\item Library books and equipment.
\item Library costs and revenue.
\item Library debt.
\item Library staff.
\item Other education.
\item Other educational costs.
\item Other library.
\item Paid educational childcare staff.
\item Paid educational staff by grade.
\item School enrollment by grade (share).
\item Student-teacher ratio by grade.
\item Student qualifications and test results.
\item Support for study organizations.
\item Teacher education by grade.
\end{itemize}

\vspace{2mm}

\item \underline{\textbf{Family Care.}}

\begin{itemize}
\itemsep 0em
\item Family care costs.
\item Family care employment.
\item Family care income.
\item Family care revenue.
\end{itemize}

\vspace{2mm}

\item \underline{\textbf{Health.}}

\begin{itemize}
\itemsep 0em
\item Health care employment.
\item Health care expenditures per person.
\end{itemize}

\vspace{2mm}

\item \underline{\textbf{Housing.}}

\begin{itemize}
\itemsep 0em
\item Housing costs.
\item Holiday houses per 1000 persons.
\item New apartments per 1000 persons.
\item New houses per 1000 persons.
\end{itemize}

\vspace{2mm}

\item \underline{\textbf{Infrastructure.}}

\begin{itemize}
\itemsep 0em
\item Air traffic costs.
\item Building structure costs.
\item Communication network costs.
\item Infrastructure revenue and cost.
\item Investment in infrastructure.
\item Physical and technical planning costs and income.
\item Port, harbor, and sea costs.
\item Road, rail, bus, and parking costs.
\end{itemize}

\vspace{2mm}

\item \underline{\textbf{Labor.}}

\begin{itemize}
\itemsep 0em
\item Gender wage gap.
\item Long-term unemployment by age.
\item Unemployment by age.
\item Wages.
\end{itemize}

\vspace{2mm}

\item \underline{\textbf{Migration.}}

\begin{itemize}
\itemsep 0em
\item Domestic occupants.
\item Domestic relocation.
\item Emigration.
\item Emigration by age range.
\item Foreign-born residents.
\item Immigration.
\item Immigration by age range.
\item Net relocation (number).
\item Population change, 1-year.
\item Population change, 5-years.
\item Refugee population and costs.
\end{itemize}

\vspace{2mm}

\item \underline{\textbf{Politics.}}

\begin{itemize}
\itemsep 0em
\item Audit costs.
\item Election turnout (county).
\item Election turnout (EU parliament).
\item Election turnout (municipal).
\item Election turnout (parliamentary).
\item Employment in politics.
\item Foreign-born politicians.
\item Political activity costs.
\item Political activity revenue.
\item Politicians by age (share).
\item Politician income.
\item Vote share for political parties.
\item Women's representation in politics.
\end{itemize}

\vspace{2mm}

\item \underline{\textbf{Public Finance.}}

\begin{itemize}
\itemsep 0em
\item Budget balance.
\item Cash flow.
\item Debt.
\item Depreciation.
\item Equity and assets.
\item Expenses on property.
\item Financial income.
\item Financial ratios.
\item Fixed assets.
\item Guarantees and liabilities.
\item Government grants.
\item Income equalization.
\item Interest expenses.
\item Investment income.
\item Net cost of municipal activities.
\item Other budget items.
\item Other finance.
\item Profits.
\item Provisions.
\item Revenue from municipal activities.
\item Self-financing rate.
\item Working capital.
\end{itemize}

\vspace{2mm}

\item \underline{\textbf{Public Service.}}

\begin{itemize}
\itemsep 0em
\item Cost equalization.
\item Cost of activities.
\item Cost of activity by category.
\item Crime rates.
\item Equalization system income.
\item Infrastructure production.
\item Net costs by category.
\item Operating expenses.
\item Paid staff.
\item Paid staff by age.
\item Paid staff by agency.
\item Paid staff by education.
\item Paid staff hours.
\item Personnel expenses (share).
\item Purchase of business by counterparty.
\item Regulatory contribution.
\item Revenue by service.
\item School rankings.
\item School resources.
\item Spending on disability assistance.
\item Spending on government activities.
\item Student performance.
\end{itemize}

\vspace{2mm}

\item \underline{\textbf{Taxes.}}

\begin{itemize}
\itemsep 0em
\item Central government taxes.
\item Compensation rate.
\item County council taxes.
\item County taxes.
\item Equalization basis.
\item Grossing-up factor.
\item Guaranteed tax power.
\item Municipal taxes.
\item Tax power.
\end{itemize}

\vspace{2mm}

\item \underline{\textbf{Utilities.}}

\begin{itemize}
\itemsep 0em
\item Communication investment.
\item Energy, water, and waste investment and costs.
\item Utility prices.
\item Working areas and premises costs.
\end{itemize}

\end{enumerate}

\clearpage

\subsection{Monte Carlo Simulation}\label{monte_carlo}

The Monte Carlo simulation exercise can be broken down into two steps: a data-generation step and an estimation step. For the data-generation step, we simulate the house price and tax setting processes shown in Equations (\ref{eqn:monte_carlo_1}) and (\ref{eqn:monte_carlo_2}).

\begin{equation} \label{eqn:monte_carlo_1}
p_{jt} = \tau_{jt} \theta_{0} + g_{0}(\zeta_{g}(x_{jt}) + \gamma_{j} + \eta_{t}) + u_{jt}
\end{equation}

\begin{equation} \label{eqn:monte_carlo_2}
\tau_{jt} = m_{0}(\zeta_{m}(x_{jt}) + \delta_{j} + \xi_{t}) + v_{jt}
\end{equation}

\noindent Since the choice of $g_{0}$ and $m_{0}$ is arbitrary, we will select linear functions. To introduce a nonlinearity into the model, we assume that $\zeta_{g}$ and $\zeta_{m}$ are both the sigmoid function, $\sigma()$, as shown in Equations (\ref{eqn:monte_carlo_3})-(\ref{eqn:monte_carlo_4}).

\begin{equation}\label{eqn:monte_carlo_3}
\zeta_{g}(x_{jt}) = \sigma(g_{0}^{0} x_{jt}^{0} + ... + g_{0}^{k-1} x_{jt}^{k-1})
\end{equation}

\begin{equation}\label{eqn:monte_carlo_4}
\zeta_{m}(x_{jt}) = \sigma(m_{0}^{0} x_{jt}^{0} + ... + m_{0}^{k-1} x_{jt}^{k-1})
\end{equation}

\noindent Furthermore, we will assume the true value of $\theta_{0}$ is -0.5. This is within the range of most of our parameter estimates in the paper, which are between 0 and -1.0; however, the choice of true value is arbitrary and should not impact the results of our simulation. 

The controls are drawn from independent normal distributions and are included in both the process that determines house prices and the process that determines tax rates. The coefficient vectors for the controls, $g$ and $m$, are drawn from a multivariate normal distribution and have a positive correlation (0.25). The fixed effects are drawn from multivariate normal distribution. They are assumed to be positively correlated (0.25) across the tax and house price processes.

We then estimate $\theta_{0}$ using OLS, DML-LASSO, and DML-DW, as described in Section \ref{Est}. For the OLS estimates, we use OLS and a random subset of the controls. Additionally, for DML-LASSO and DML-DW, we include the full set of controls.

We perform the simulation with 947 controls, 7 time periods, and 290 geographic entities. This is in line with the sample we use in most regressions in the paper. We repeat the process 100 times, subtract the true value from the estimated values, and then and plot the distribution of coefficient estimate biases in Figure \ref{fig:monte_carlo}. 
\end{document}